\documentclass[amsmath,amssymb,aps,prd,11pt,tightenlines,superscriptaddress,nofootinbib,preprintnumbers,notitlepage]{revtex4-1}

\usepackage{wrapfig}

\input{header}

\usepackage{datetime}

\usepackage{lineno}

\begin{document}

\pagestyle{empty}

\vspace*{-0.1in}

\vspace*{-0.1in}
\title{{\large SCIENCE AND PROJECT PLANNING FOR THE FORWARD PHYSICS FACILITY IN PREPARATION FOR THE 2024--2026 EUROPEAN PARTICLE PHYSICS STRATEGY UPDATE} \\
\vspace*{1.68in} 
{\mbox{\normalsize \normalfont on behalf of the FPF Working Groups}} 
\vspace*{-1.64in} 
}

\author{Jyotismita Adhikary}
\affiliation{National Centre for Nuclear Research, Pasteura 7, Warsaw, PL-02-093, Poland}

\author{Luis~A.~Anchordoqui}
\affiliation{Department of Physics and Astronomy, Lehman College, City University of New York, Bronx, NY 10468, USA}

\author{Akitaka Ariga}
\affiliation{Albert Einstein Center for Fundamental Physics, Laboratory for High Energy Physics, \\ University of Bern, Sidlerstrasse 5, CH-3012 Bern, Switzerland}
\affiliation{\mbox{Department of Physics, Chiba University, 1-33 Yayoi-cho Inage-ku, Chiba, 263-8522, Japan}}

\author{Tomoko Ariga}
\affiliation{Kyushu University, Nishi-ku, 819-0395 Fukuoka, Japan}

\author{Alan~J.~Barr}
\affiliation{Department of Physics, University of Oxford, OX1 3RH, United Kingdom}

\author{Brian Batell}
\affiliation{Department of Physics and Astronomy, University of Pittsburgh, Pittsburgh, PA 15217, USA}

\author{Jianming Bian}
\affiliation{Department of Physics and Astronomy, University of California, Irvine, CA 92697-4575, USA}

\author{Jamie Boyd}
\affiliation{CERN, CH-1211 Geneva 23, Switzerland}

\author{Matthew Citron}
\affiliation{\mbox{Department of Physics and Astronomy, University of California, Davis, CA 95616, USA}}

\author{Albert De Roeck}
\affiliation{CERN, CH-1211 Geneva 23, Switzerland}

\author{Milind~V.~Diwan}
\affiliation{Brookhaven National Laboratory, Upton, NY 11973, USA}

\author{Jonathan~L.~Feng}
\affiliation{Department of Physics and Astronomy, University of California, Irvine, CA 92697-4575, USA}

\author{Christopher~S.~Hill}
\affiliation{Department of Physics, The Ohio State University, Columbus, OH 43210, USA}

\author{Yu Seon Jeong}
\affiliation{\mbox{Department of Physics and Astronomy, University of Iowa, Iowa City, IA 52246, USA}}

\author{Felix Kling}
\affiliation{Deutsches Elektronen-Synchrotron DESY, Notkestr. 85, 22607 Hamburg, Germany}

\author{Steven Linden}
\affiliation{Brookhaven National Laboratory, Upton, NY 11973, USA}

\author{Toni~M\"akel\"a}
\affiliation{Department of Physics and Astronomy, University of California, Irvine, CA 92697-4575, USA}

\author{Kostas Mavrokoridis}
\affiliation{University of Liverpool, Liverpool L69 3BX, United Kingdom}

\author{Josh McFayden}
\affiliation{Department of Physics \& Astronomy, University of Sussex, Sussex House, Falmer, Brighton, BN1 9RH, United Kingdom}

\author{Hidetoshi Otono}
\affiliation{Kyushu University, Nishi-ku, 819-0395 Fukuoka, Japan}

\author{Juan Rojo}
\affiliation{\mbox{Department of Physics and Astronomy, VU Amsterdam, 1081 HV Amsterdam, The Netherlands}}
\affiliation{\mbox{Nikhef Theory Group, Science Park 105, 1098 XG Amsterdam, The Netherlands}}

\author{Dennis Soldin}
\affiliation{\mbox{Department of Physics and Astronomy, University of Utah, Salt Lake City, UT 84112, USA}}

\author{Anna Stasto}
\affiliation{Department of Physics, Penn State University, University Park, PA 16802, USA}

\author{Sebastian Trojanowski}
\affiliation{National Centre for Nuclear Research, Pasteura 7, Warsaw, PL-02-093, Poland}

\author{Matteo Vicenzi}
\affiliation{Brookhaven National Laboratory, Upton, NY 11973, USA} 

\author{Wenjie Wu\vspace{0.5in}}
\affiliation{Department of Physics and Astronomy, University of California, Irvine, CA 92697-4575, USA}

\begin{abstract}
\vspace*{0.2in}
The recent direct detection of neutrinos at the LHC has opened a new window on high-energy particle physics and highlighted the potential of forward physics for groundbreaking discoveries.  In the last year, the physics case for forward physics has continued to grow, and there has been extensive work on defining the Forward Physics Facility and its experiments to realize this physics potential in a timely and cost-effective manner.  Following a 2-page Executive Summary, we present the status of the FPF, beginning with the FPF's unique potential to shed light on dark matter, new particles, neutrino physics, QCD, and astroparticle physics.  We summarize the current designs for the Facility and its experiments, FASER2, FASER$\nu$2, FORMOSA, and FLArE, and conclude by discussing international partnerships and organization, and the FPF's schedule, budget, and technical coordination.
\end{abstract}

\maketitle
\clearpage

\section*{Executive Summary
\label{sec:executivesummary}}

High-energy colliders have enabled many groundbreaking discoveries since they were first constructed over 60 years ago.  As the latest example, the Large Hadron Collider (LHC) at CERN has been the center of attention in particle physics for decades. 
Despite this, the physics potential of the LHC is far from being fully explored, because the large detectors at the LHC are blind to collisions that produce particles in the forward direction, along the beamline. 
These forward collisions are a treasure trove of physics, providing the only way to study TeV neutrinos produced in the lab and unique opportunities to discover and study dark matter and other new particles beyond the Standard Model of particle physics.  

\begin{wrapfigure}{R}{0.40\textwidth}
\vspace*{-0.4in}
  \begin{center} \includegraphics[width=0.40\textwidth]{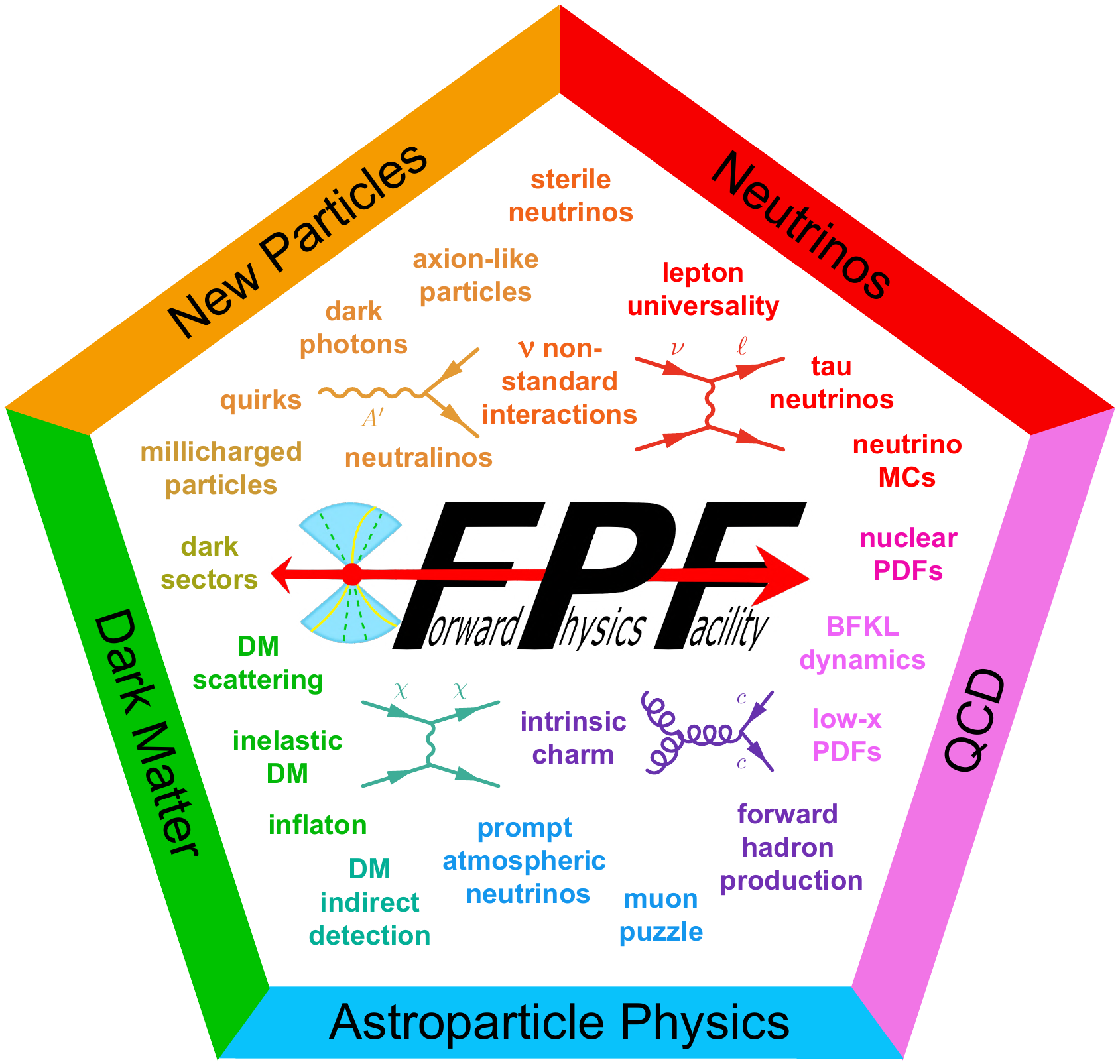}
\caption{The rich physics program at the FPF spans many topics and frontiers.
\label{fig:pentagon}}
  \end{center}
\vspace*{-0.3in}
\end{wrapfigure}

The Forward Physics Facility (FPF) is a proposal to build a new underground cavern at CERN to house a suite of forward experiments during the High-Luminosity LHC (HL-LHC) era.
These experiments will cover the blind spots of the existing LHC detectors and are required if the LHC is to fully realize its physics potential. 
The physics program of the FPF is broad and deep; see \cref{fig:pentagon}.  The FPF can discover a wide variety of new particles that cannot be discovered at fixed target facilities or other LHC experiments. 
In the event of a discovery, the FPF, with other experiments, will play an essential role in determining the precise nature of the new physics and its possible connection to the dark universe.  In addition, the FPF is the only facility that will be able to detect millions of neutrinos with TeV energies, enabling precision probes of neutrino properties for all three flavors. 
These neutrinos will also sharpen our understanding of proton and nuclear structure, enhancing the power of new particle searches at ATLAS and CMS, and enabling IceCube, Auger, KM3NeT and other astroparticle experiments to make the most of the new era of multi-messenger astronomy. 

\begin{figure}[tbp]
\centering
\includegraphics[width=0.97\textwidth]{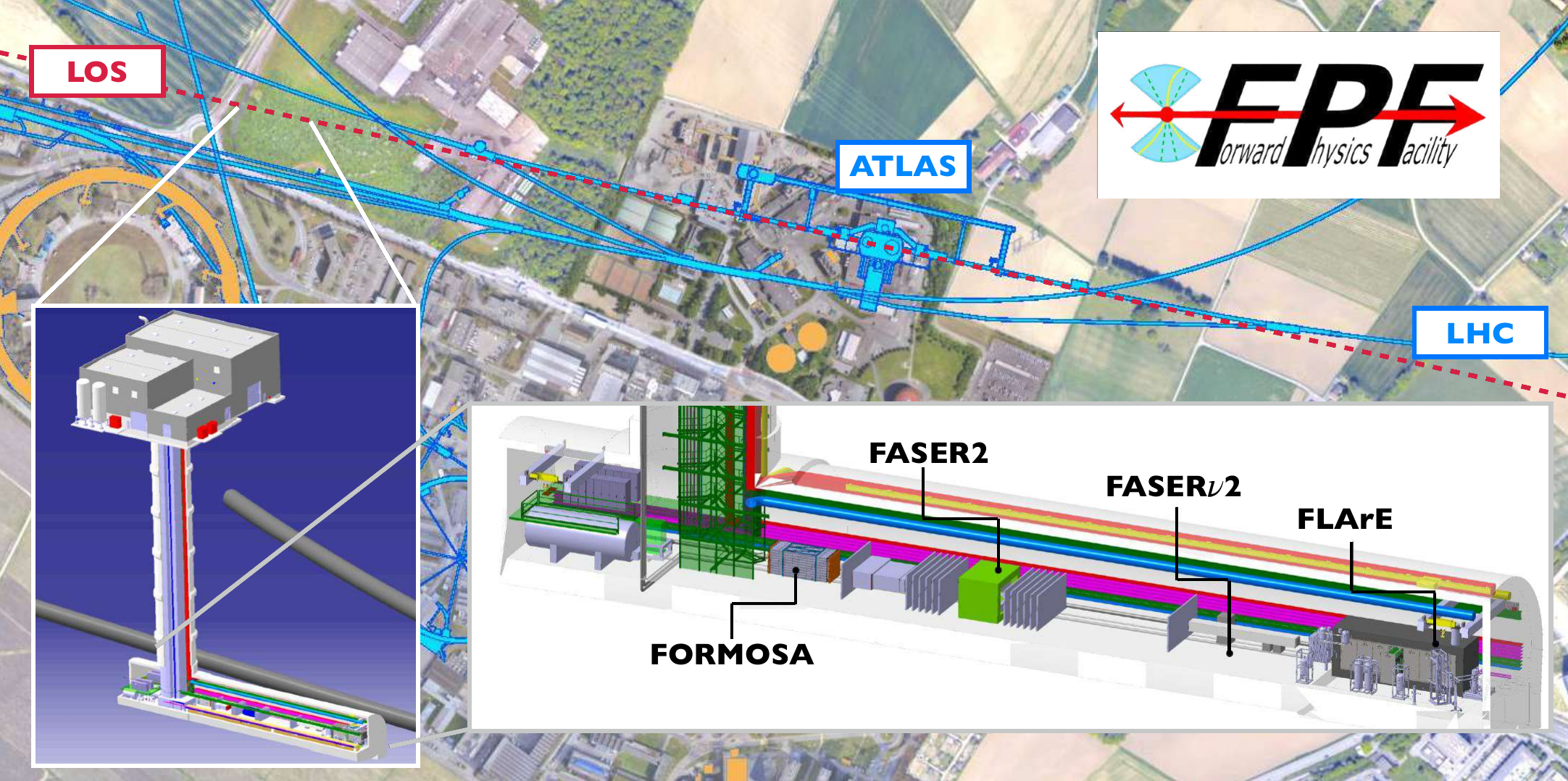}
\caption{The FPF is located 627--702 m west of the ATLAS IP along the line of sight. The FPF cavern is 75~m long and 12~m wide and will house a diverse set of experiments to fully explore the forward region. \vspace*{-0.05in}}
\label{fig:ExecutiveSummaryMap}
\end{figure}

\vspace*{0.08in} 
\noindent {\bf The Facility:}
An extensive site selection study has been conducted by the CERN Civil Engineering group.  The resulting site is shown in \cref{fig:ExecutiveSummaryMap}. This location is shielded from the ATLAS interaction point (IP) by over 200 m of concrete and rock, providing an ideal location to search for rare processes and very weakly interacting particles.  Vibration, radiation, and safety studies have shown that the FPF can be constructed independently of the LHC without interfering with LHC operations.  A core sample, taken along the location of the 88 m-deep shaft to provide information about the geological conditions, has confirmed that the site is suitable for construction.
Studies of LHC-generated radiation have concluded that the facility can be safely accessed with appropriate controls during beam operations.  Flexible, safe access will allow the construction and operation of FPF experiments to be fully independent of the LHC, greatly simplifying schedules and budgets.
In fact, experiments can be constructed and modified in the cavern while the LHC beam is on and other experiments are taking data, allowing a flexible program that can respond quickly to the latest developments and discoveries in particle physics and beyond.  

\vspace*{0.08in} 
\noindent {\bf The Experiments:}
The FPF is uniquely suited to explore physics in the forward region because it will house a diverse set of experiments based on different detector technologies and optimized for particular physics goals. The proposed experiments are shown in \cref{fig:ExecutiveSummaryMap} and include
\begin{itemize}
\setlength\itemsep{-0.05in}
\item FASER2, a magnetic tracking spectrometer, designed to search for light and weakly-interacting states, including new force carriers, sterile neutrinos, axion-like particles, and dark sector particles, and to distinguish $\nu$ and $\bar{\nu}$ charged current scattering in the upstream detectors. 
\item FASER$\nu$2, an on-axis emulsion detector, with pseudorapidity range $\eta > 8.4$, that will detect $\sim 10^6$ neutrinos at TeV energies with unparalleled spatial resolution, including several thousands of tau neutrinos, among the least studied of all the known particles.
\item FLArE, a 10-ton-scale, noble liquid, fine-grained time projection chamber that will detect neutrinos and search for light dark matter with high kinematic resolution, wide dynamic range and good particle-identification capabilities. 
\item FORMOSA, a detector composed of scintillating bars, with world-leading sensitivity to millicharged particles across a large range of masses.
\end{itemize}
\vspace*{-0.1in}

\vspace*{0.08in} 
\noindent {\bf Cost and Timeline:}
All of the planned experiments are relatively small, low cost, require limited R\&D, and can be constructed in a timely way. A Class 4 cost estimate\footnote{According to international standards of conventional construction, a Class 4 estimate has a range of -30$\%$ to +50$\%$ around the point estimate~\cite{CEcosting}.}
for the Facility by the CERN engineering and technical teams is 35 MCHF for the construction of the new shaft and cavern and $\mathcal{O}$(10MCHF) for the installation of infrastructure and services.  Cost estimates for the experiments range from 2 MCHF for FORMOSA to 15 MCHF for FASER$\nu$2, as detailed in this report. The FPF requires no modifications to the LHC and will support a sustainable experimental program, without additional power consumption for the beam beyond the existing LHC program.  

To fully exploit the forward physics opportunities, which will disappear for several decades if not explored in the 2030s, the FPF and its experiments should be ready for physics in the HL-LHC era as early as possible in Run 4.  A possible timeline is for the FPF to be built during Long Shutdown 3 from 2026-29, the support services and experiments to be installed starting in 2029, and the experiments to begin taking data during Run 4. All of the experiments will be supported by international collaborations, and, as the physics program begins in LHC Run 4 from 2030-33, after HL-LHC upgrades are completed, the FPF will attract a large and diverse global community.
In addition, as a mid-scale project composed of smaller experiments that can be realized on short and flexible timescales, the FPF will provide a multitude of scientific and leadership opportunities for junior researchers, who can make important contributions from construction to data analysis in a single graduate student lifetime.
Such a timeline is guaranteed to produce novel physics results through studies of very high-energy neutrinos, QCD, and other topics, and will additionally enhance the HL-LHC's potential for groundbreaking discoveries for many years to come.

\section{Physics at the FPF
\label{sec:physics}}

The science case for the FPF has been developed in several dedicated FPF meetings~\cite{FPF1Meeting, FPF2Meeting, FPF3Meeting, FPF4Meeting, FPF5Meeting, FPF6Meeting, FPF7Meeting, FPFTheoryWS}. 
The opportunities have been summarized in an 80-page review~\cite{Anchordoqui:2021ghd} and a more comprehensive 430-page White Paper~\cite{Feng:2022inv}, written and endorsed by 400 physicists.  The current paper extends and refines the discussion presented in Refs.~\cite{Anchordoqui:2021ghd, Feng:2022inv}.

\begin{figure}[b]
\centering
\includegraphics[width=0.95\textwidth]{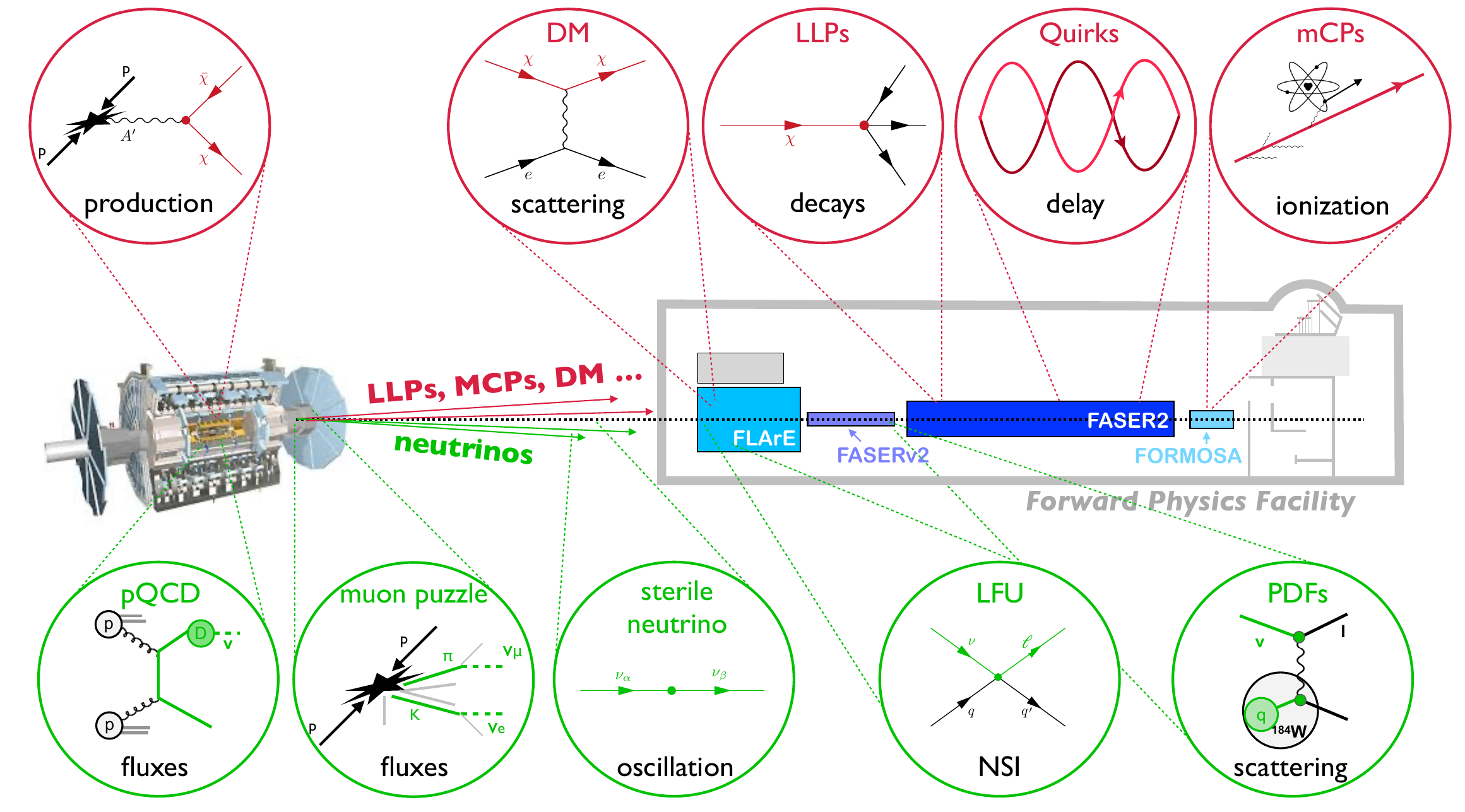}
\caption{\textbf{New particle searches and neutrino measurements at the FPF.} 
Representative examples of DM and other new particles that can be discovered and studied at the FPF (top) and of some of the many topics that will be illuminated by TeV-energy neutrino measurements at the FPF (bottom). }
\label{fig:BSM_overview}
\end{figure}

As illustrated in \cref{fig:BSM_overview}, the FPF physics program encompasses a broad set of searches for novel new physics and unique Standard Model (SM) measurements that leverage the diverse capabilities of the suite of FPF experiments. On the side of new physics searches, this includes long-lived-particle decays to visible final states that are being probed at FASER~\cite{FASER:2018ceo}, dark matter (DM) scattering signatures that can be probed at FLArE~\cite{Batell:2021blf}, and unconventional ionization caused by new particles with fractional electric charge, which can be seen at FORMOSA~\cite{Foroughi-Abari:2020qar}. 
The SM measurements leverage the unprecedented flux of collider neutrinos that can be observed by FLArE and FASER$\nu$2 to study lepton flavor universality and non-standard interactions in the neutrino sector, probe QCD dynamics in novel kinematic regions, and resolve outstanding conundrums in astroparticle physics. 

In the following, we present a few highlights of this broad program.  Comprehensive discussions of the physics potential of the FPF can be found in Refs.~\cite{Feng:2022inv, Anchordoqui:2021ghd}. 

\subsection{Dark Matter
\label{sec:science_dm} }

The DM puzzle stands out as one of the foremost motivations for beyond-the-SM (BSM) physics. The form of DM realized in nature is unknown, and there are well-motivated possibilities that can only be probed by experiments at the FPF.

A generic and compelling possibility is that DM is part of a dark sector, feebly coupled to the SM by a mediator particle via a portal interaction. In this scenario, the DM relic abundance can be produced through simple thermal freeze-out, extending the traditional WIMP production mechanism to DM masses in the MeV to GeV range. Such light DM, along with the associated mediator particles and other dark sector states, can be copiously produced in the forward region at the LHC, thus providing key experimental targets for BSM searches at the FPF. The search strategies required to probe these scenarios at the FPF depend on the structure of the dark sector and its mass spectrum, and we highlight a few possibilities in the following. \medskip

\begin{figure}[b]
\centering
\includegraphics[width=0.49\textwidth]{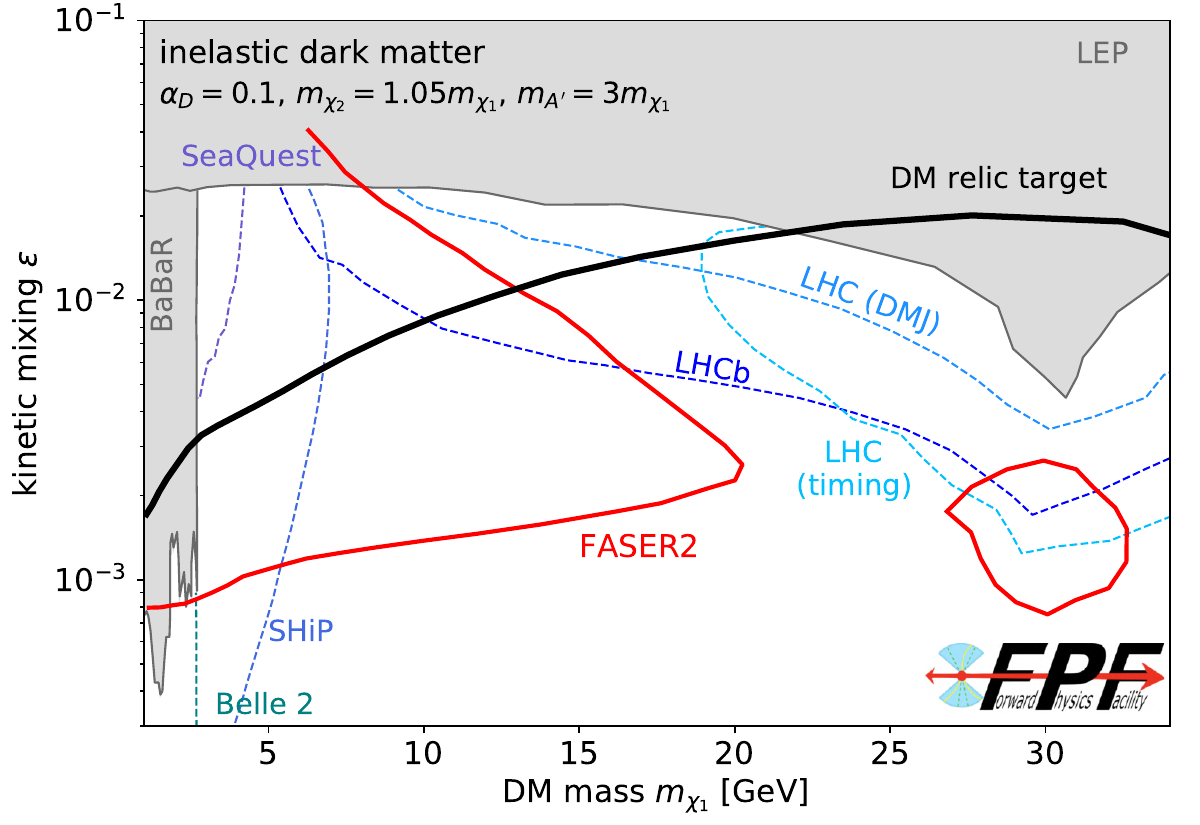}
\includegraphics[width=0.49\textwidth]{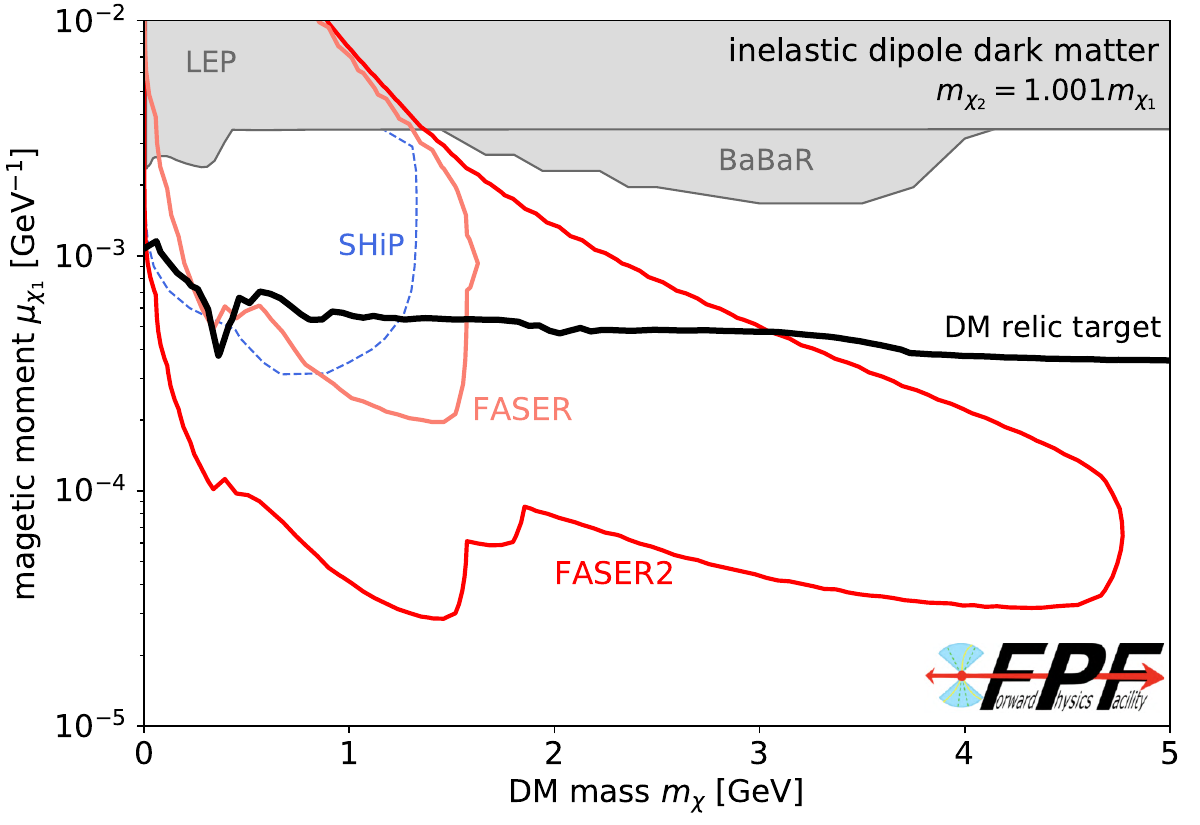}
\caption{\textbf{Inelastic dark matter searches at the FPF}. The discovery potential of FASER2 and other experiments for two different realizations of inelastic DM. The left panel considers a the case of heavy inelastic DM interacting via a dark photon portal, as introduced in ~\cite{Berlin:2018jbm}, where the high energy of the LHC allows FASER2 to probe masses up to tens of GeV.  The right panel considers the case of light inelastic DM with very small mass splittings that is mediated by a dipole portal as introduced in Ref.~\cite{Dienes:2023uve}, where the large LHC energy boosts the signal to observable energies. In both scenarios, the reach of FASER2 extends beyond all other experiments, including direct and indirect DM searches, LHC experiments, and beam dump experiments, such as SHiP. It also covers the thermal DM relic target (solid black lines), that is the cosmologically-favored parameter space where the model predicts the observed dark matter relic abundance as produced through thermal freeze-out. For comparison, we have shown the leading constraints provided by  BaBaR~\cite{BaBar:2017tiz} and LEP~\cite{Hook:2010tw,Curtin:2014cca}, as well as projections from a number of other proposed searches, including those for displaced muon jets (DMJ) and delayed particles (timing) at the main LHC experiments~\cite{Izaguirre:2015zva, Liu:2018wte} as well as displaced particle searches at LHCb~\cite{LHCb:2017trq,Ilten:2016tkc,Pierce:2017taw}, SHiP~\cite{SHiP:2021nfo}, Belle 2~\cite{Belle-II:2018jsg}, and SeaQuest~\cite{Berlin:2018pwi}.}
\label{fig:BSM_DM}
\end{figure}

\noindent \textbf{Mediators to DM:}
If the mediator is the lightest state in the dark sector, it will decay back to SM particles through the portal interaction. Due to the feebleness of this coupling, the mediators can easily possess a macroscopic decay length, thus manifesting at the FPF as a visibly-decaying long-lived particle. The powerful capability of the FPF to search for a broad spectrum of long-lived particles has been established in a large number of publications and summarized in Ref.~\cite{Feng:2022inv}. Notably, this includes all of the benchmarks models discussed in the context of the Physics Beyond Colliders initiative~\cite{Alemany:2019vsk}: dark photons, dark Higgs, heavy neutral leptons, and axion-like particles.  It is worth emphasizing that in the event of a long-lived particle discovery, multiple experiments with complementary experimental capabilities will be required to determine the fundamental properties of the new state (i.e., its mass, lifetime, spin, and couplings) and its possible connection to the dark universe, and the FPF experiments will play an essential role in this endeavor. 
\medskip 

\noindent \textbf{Inelastic DM:} 
In addition, there are also well-motivated DM scenarios featuring a rich dark sector structure that can be uniquely probed at the FPF. This is nicely illustrated in \cref{fig:BSM_DM}, which shows the expected sensitivity of FASER2 to two realisations of inelastic DM (iDM). This model contains an excited dark sector state that decays into a somewhat lighter DM particle plus a visible final state. The left panel considers a relatively heavy iDM scenario with masses in the tens of GeV range~\cite{Berlin:2018jbm}. Such states are beyond the kinematic threshold of beam dump experiments, but the high energies available at the LHC imply significant production rates, and the sensitivity of the FPF to highly-displaced decays allows it to uniquely explore new regions of parameter space beyond the reach of the existing large LHC detectors. The right panel considers a case with a very small mass splitting between the excited state and the DM~\cite{Dienes:2023uve}. Due to the large particle energies in the forward direction of the LHC, sufficiently energetic signals can be observed the FPF, while a corresponding signal at beam dump experiments would be below the threshold of detectability.
In both scenarios FASER2 will be able to decisively test a broad swath of parameter space where DM is produced in the early universe through thermal freeze-out. \medskip

\noindent \textbf{DM Scattering:}
Another scenario of interest arises when the DM is significantly lighter than the mediator. In this case, the mediators produced in LHC proton collisions decay ``invisibly'' to pairs of DM particles, resulting in a significant flux of DM particles directed in the forward region at the LHC. Such DM can then be detected through its scattering with electrons and nuclei at FPF detectors, such as FLArE and FASER$\nu$2~\cite{Batell:2021blf, Batell:2021aja, Batell:2021snh}. In simple dark sector models with a dark photon or hadrophilic vector mediator, these experiments will be able to probe new regions of parameter space that are compatible with a thermally-produced DM relic abundance. The ability of FLArE and FASER$\nu$2 to detect DM scattering in the relativistic regime offers an experimental probe that complements traditional underground DM direct detection (DD) experiments. Notably, the expected signal rates in accelerator-based searches exhibit a different dependence on DM interaction strength compared to DD experiments, assuming thermal DM production in the early Universe. Consequently, simultaneous discovery in both experimental approaches would enable better discrimination between various DM scenarios. Additionally, the FPF searches provide insights into DM interactions characterized by suppressed non-relativistic scattering rates, which are otherwise challenging to probe. These proposed searches, based on the direct observation of DM scattering, complement dedicated missing energy/momentum accelerator-based experiments. 

\subsection{New Particles 
\label{sec:science_bsm} }

\begin{figure}[thb]
\centering
\includegraphics[width=0.49\textwidth]{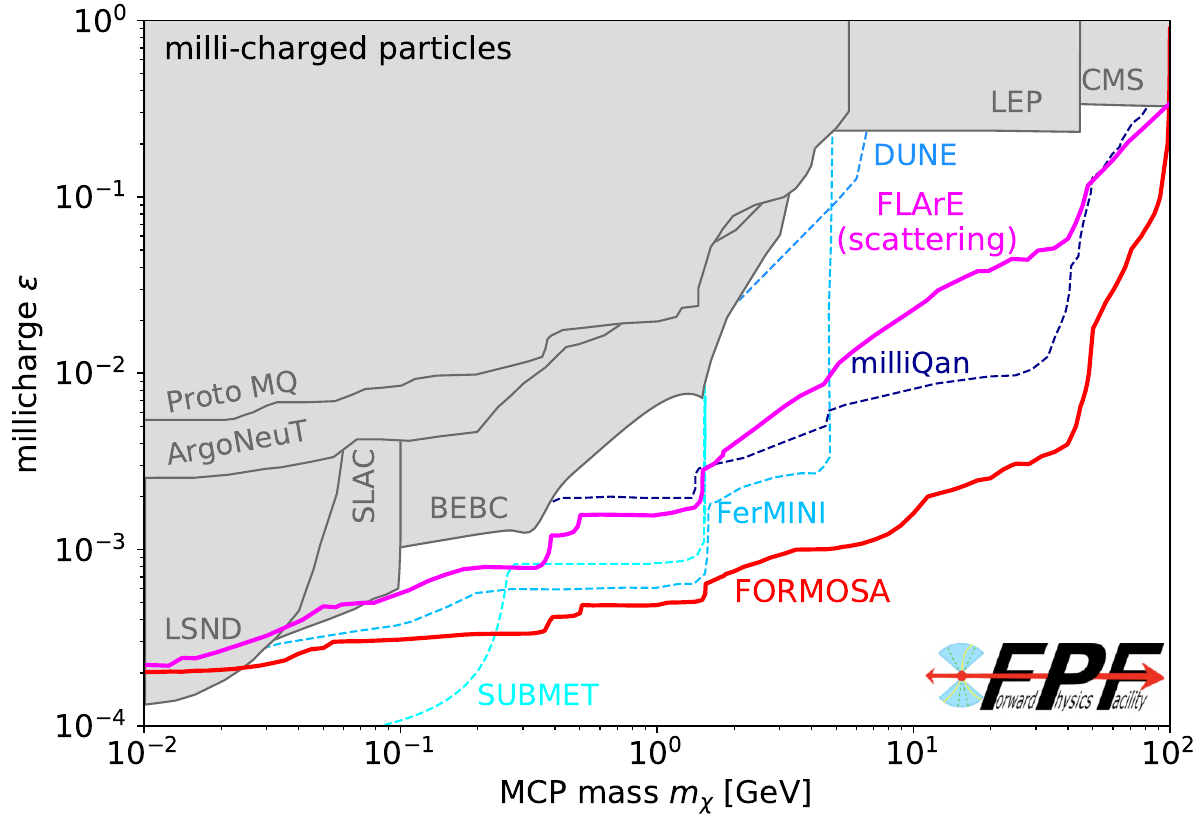}
\hfill
\includegraphics[width=0.49\textwidth]{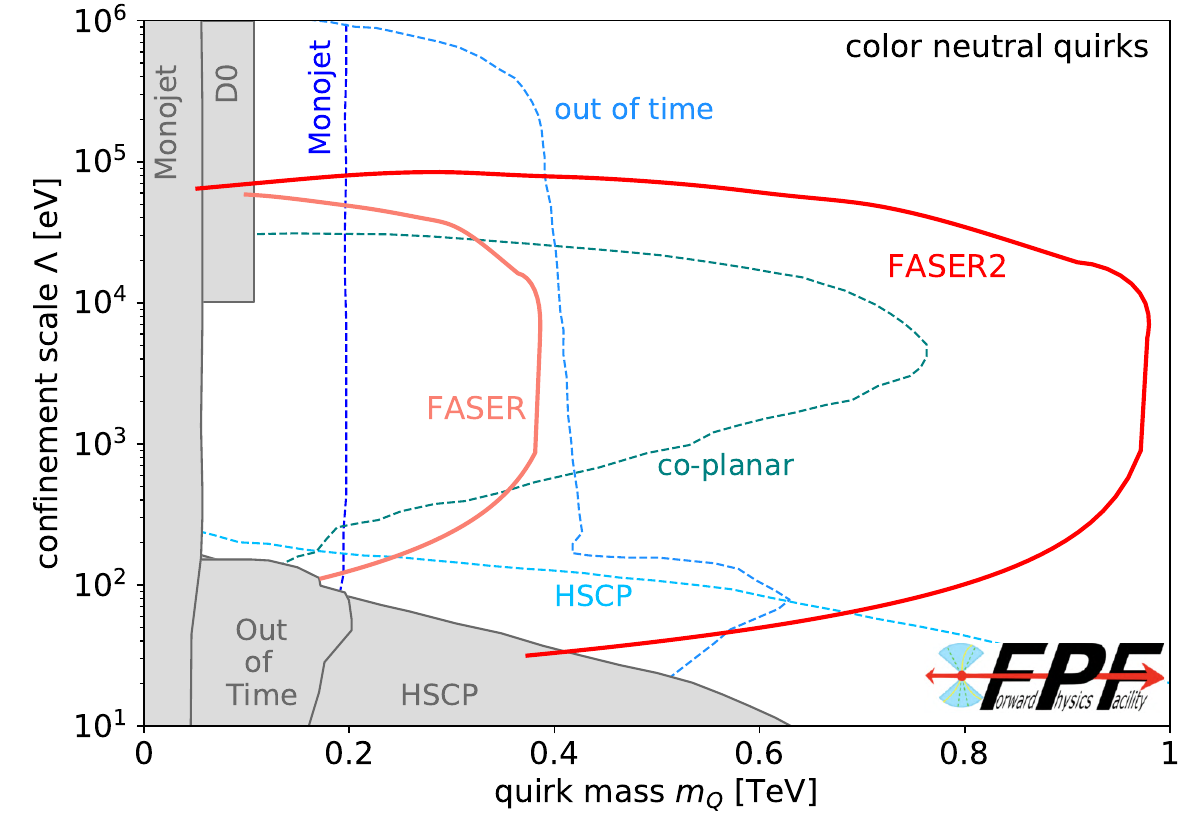}
\caption{\textbf{New particle searches at the FPF.} Left: The discovery reach of FORMOSA and FLArE for millicharged particles~\cite{Foroughi-Abari:2020qar, Kling:2022ykt}. Right: The discovery reach of FASER and FASER2 for color-neutral quirks~\cite{Feng:2024zgp}.  In both panels, we also show existing bounds (gray shaded regions) and projected sensitivities of other experiments (dashed contours), including BEBC~\cite{Marocco:2020dqu}, SLAC~\cite{Prinz:1998ua}, LEP~\cite{Davidson:2000hf,OPAL:1995uwx}, CMS~\cite{CMS:2012xi, Jaeckel:2012yz}, LSND~\cite{Magill:2018tbb}, ArgoNeuT~\cite{ArgoNeuT:2019ckq}, Proto-milliQan~\cite{Ball:2020dnx}, milliQan~\cite{Ball:2016zrp}, FerMINI~\cite{Kelly:2018brz}, SUBMET~\cite{Choi:2020mbk}, monojet searches~\cite{Farina:2017cts,CMS-PAS-EXO-16-037,ATLAS:2016bek}, quirk searches at D0~\cite{D0:2010kkd}, heavy stable charged particle searches (HSCP)~\cite{Farina:2017cts,CMS-PAS-EXO-16-036,ATLAS:2016onr}, co-planar hits searches~\cite{Knapen:2017kly}, and out-of-time searches~\cite{Evans:2018jmd,ATLAS:2013whh}.}
\label{fig:BSM_newparticles}
\end{figure}

The many experimental signatures and broad range of BSM particle masses that can be probed at the FPF, from MeV up to the TeV scale, provide the foundation for a broad BSM physics program that will address fundamental questions in particle physics in a manner that is complementary to other existing and proposed facilities. 
This includes searches for decays of long-lived particles which were discussed in Ref.~\cite{Feng:2022inv}. 
Such particles, for example, arise in models proposed to explain the nature and observed abundance of dark matter, such as dark matter mediators and inelastic dark matter discussed above; as light relaxions to solve the electroweak hierarchy problem~\cite{Banerjee:2020kww}; as light inflaton candidate~\cite{Okada:2019opp}; in the two Higgs doublet model~\cite{Kling:2022uzy} or R-partity violating models of supersymmetry~\cite{Dreiner:2022swd}. 
Further signatures are provided by muon-philic particles that can be produced in muon scattering inside the neutrino detectors~\cite{Ariga:2023fjg, Batell:2024cdl, MammenAbraham:2025gai} or light axions that can convert to photons in the magnetic fields of FPF experiments~\cite{Kling:2022ehv}.    
Below, we illustrate two additional examples of such unique search opportunities. \medskip 

\noindent \textbf{Millicharged Particles:} 
The prospects for millicharged particle (mCP) searches at the FPF are shown in the left panel of \cref{fig:BSM_newparticles}. 
Such particles provide an interesting BSM physics target, both for their possible implications for the principle of charge quantization and as a candidate for a strongly interacting sub-component of DM. FORMOSA, a proposed scintillator-based experiment at the FPF, will have world-leading sensitivity to mCPs~\cite{Foroughi-Abari:2020qar}. When compared to existing bounds and projections from several other ongoing or proposed experiments, FORMOSA benefits from the high-energy LHC collisions and the enhanced mCP production in the forward region, enabling the most sensitive probe of mCPs in the broad mass range from 100 MeV to 100 GeV.\medskip

\noindent \textbf{Quirks:} 
Quirks ($\cal{Q}$) are new particles that are charged under both the SM and an additional strongly-interacting gauge force.  At colliders, color-neutral quirks are produced at colliders through Drell-Yan production, and colored quirks are dominantly produced by processes with an $s$-channel gluon.  After they are produced at a collider, $\cal{Q} \bar{\cal{Q}}$ pairs then travel together down the beamline.  Currently, quirks are constrained by searches for heavy, stable charged particles, monojets, and other exotic signatures; for more details, see Ref.~\cite{Feng:2024zgp}.  Discovery prospects for quirks are shown in the right panel of \cref{fig:BSM_newparticles}.  For hidden confinement scales $\Lambda \agt 100~\text{eV}$, current bounds do not even exclude quirk masses of 100 GeV. At the same time, simply by searching for simultaneous pairs of slow or delayed charged tracks~\cite{Feng:2024zgp}, FASER2 will probe masses up to 1 TeV, a range motivated by neutral naturalness solutions to the gauge hierarchy problem~\cite{Batell:2022pzc}. Such heavy quirks cannot be produced in fixed-target experiments and demonstrate another unique search capability of forward detectors at high-energy colliders. 

\subsection{Neutrino Physics 
\label{sec:science_neutrino}}

\begin{figure}[thb]
\centering
\includegraphics[width=0.325\textwidth]{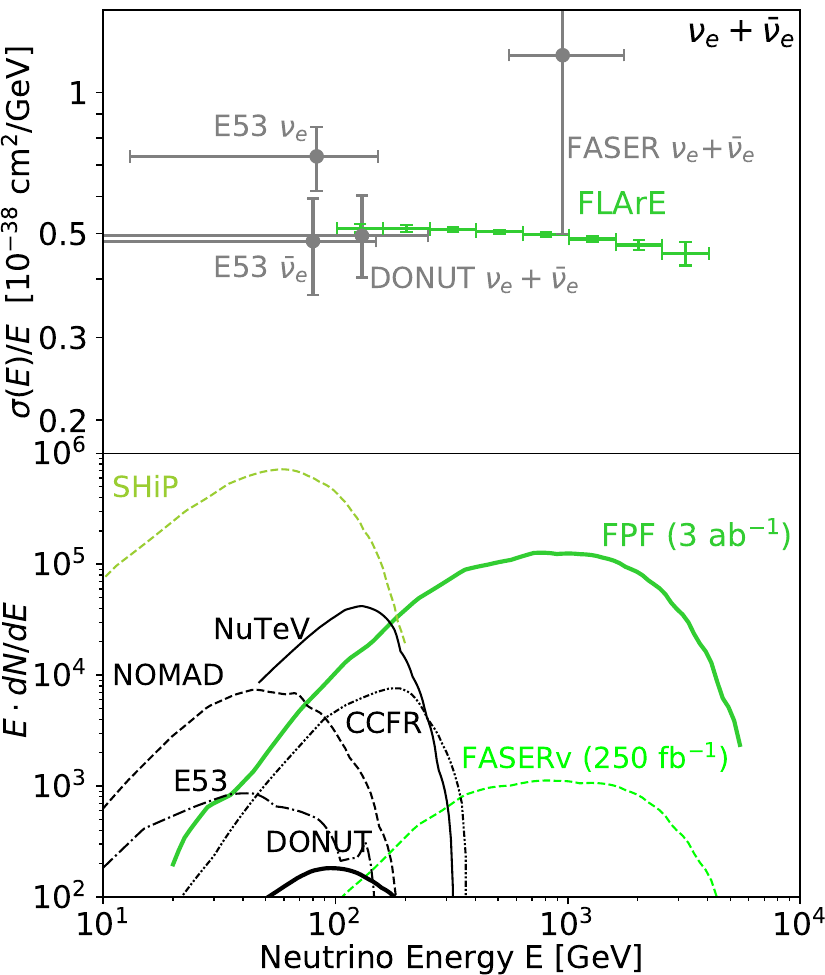}
\includegraphics[width=0.325\textwidth]{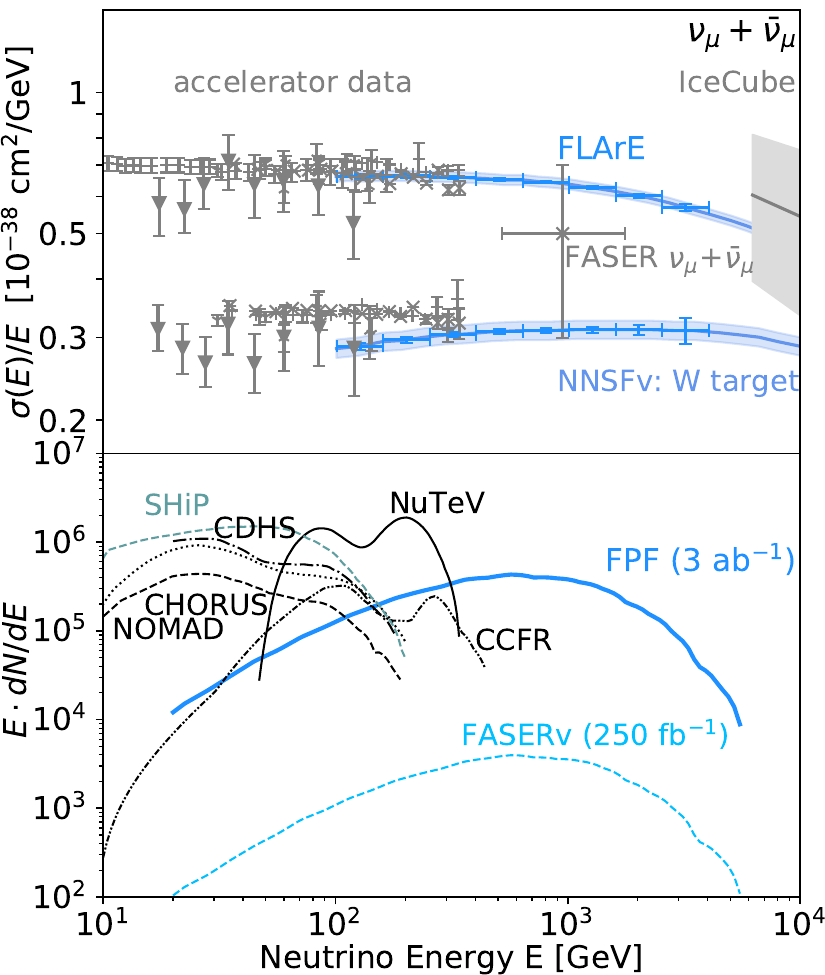}
\includegraphics[width=0.325\textwidth]{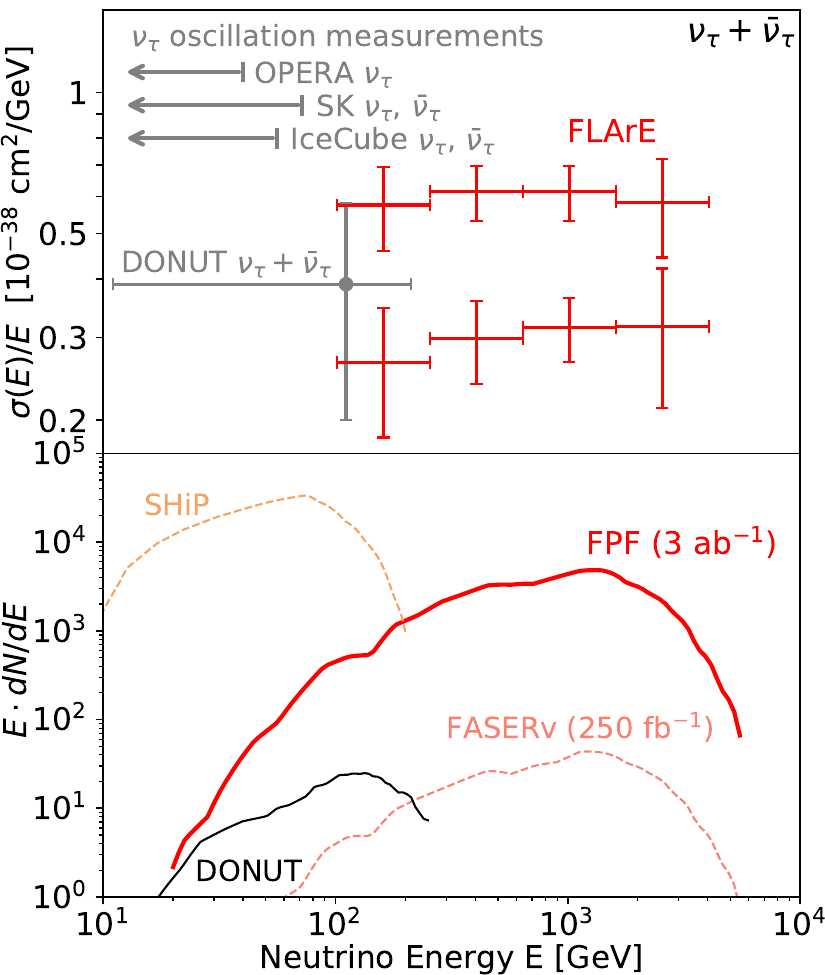}
\caption{\textbf{Neutrino yields and cross sections at the FPF.} 
The expected precision of FLArE measurements of neutrino interaction cross sections (top, statistical errors only) and the combined spectrum of neutrinos interacting the FPF experiments (bottom) as a function of energy for electron (left), muon (middle), and tau (right) neutrinos.
In the case of muon and tau neutrinos, separate measurements of the neutrino and anti-neutrino measurement can be performed using muons passing through the FASER2 spectrometer, where a 17\% branching fraction of taus into muons was considered. 
Existing data from accelerator experiments~\cite{ParticleDataGroup:2020ssz}, IceCube~\cite{IceCube:2017roe}, and the recent FASER$\nu$ result~\cite{FASER:2024hoe} are also shown, together with the prospects for SHiP.}
\label{fig:SM_neutrino}
\end{figure}

The LHC is the highest-energy particle collider built to date, and it is therefore also the source of the most energetic neutrinos produced in a controlled laboratory environment.
Indeed, the LHC generates intense, strongly collimated, and highly energetic beams of both neutrinos and anti-neutrinos of all three flavors in the forward direction.
Although this has been known since the 1980s~\cite{DeRujula:1984pg}, only recently have two detectors, FASER$\nu$~\cite{FASER:2020gpr} and SND@LHC~\cite{SNDLHC:2022ihg}, been installed to take advantage of this opportunity. These pathfinder experiments have just recently directly observed collider neutrinos for the first time~\cite{moriondresultsFASER, moriondresultsSND, FASER:2023zcr}. By the end of LHC Run 3 in 2026, these experiments are expected to detect approximately $10^4$ neutrinos. The FPF experiments, with larger detectors and higher luminosities, are projected to detect $10^5$ electron neutrino, $10^6$ muon neutrino, and $10^4$ tau neutrino interactions, providing approximately 100 times more statistics over the current experiments, enabling precision measurements for all three flavors, and distinguishing tau neutrinos from anti-neutrinos for the first time. \medskip 

\noindent \textbf{Neutrino Event Rates:}  \cref{fig:SM_neutrino} (top) displays the expected precision of FPF measurements of the neutrino-nucleon charged-current scattering cross sections for all three neutrino flavors. The low-energy region has been well-constrained by neutrino experiments using existing accelerators~\cite{ParticleDataGroup:2020ssz}. IceCube has also placed constraints on the muon neutrino cross section at very high energies using atmospheric neutrinos, although with relatively large uncertainties~\cite{IceCube:2017roe}. The bottom panels show the expected energy spectra of interacting neutrinos at the FPF, as estimated using EPOS-LHC~\cite{Pierog:2013ria} to simulate light hadrons and POWHEG matched with Pythia~\cite{Buonocore:2023kna} to simulate charm hadron production and the fast neutrino flux simulation~\cite{Kling:2021gos} to obtain the neutrino spectrum. The collider neutrino energy spectrum peaks at $\sim \tev$ energies, where currently no measurements exist. 
We also display the expected neutrino fluxes at SHiP, which peak at much lower ($\lesssim$ 100 GeV) energies.
\medskip

\begin{wrapfigure}{t}{0.50\textwidth}
  \centering  \includegraphics[width=0.50\textwidth]{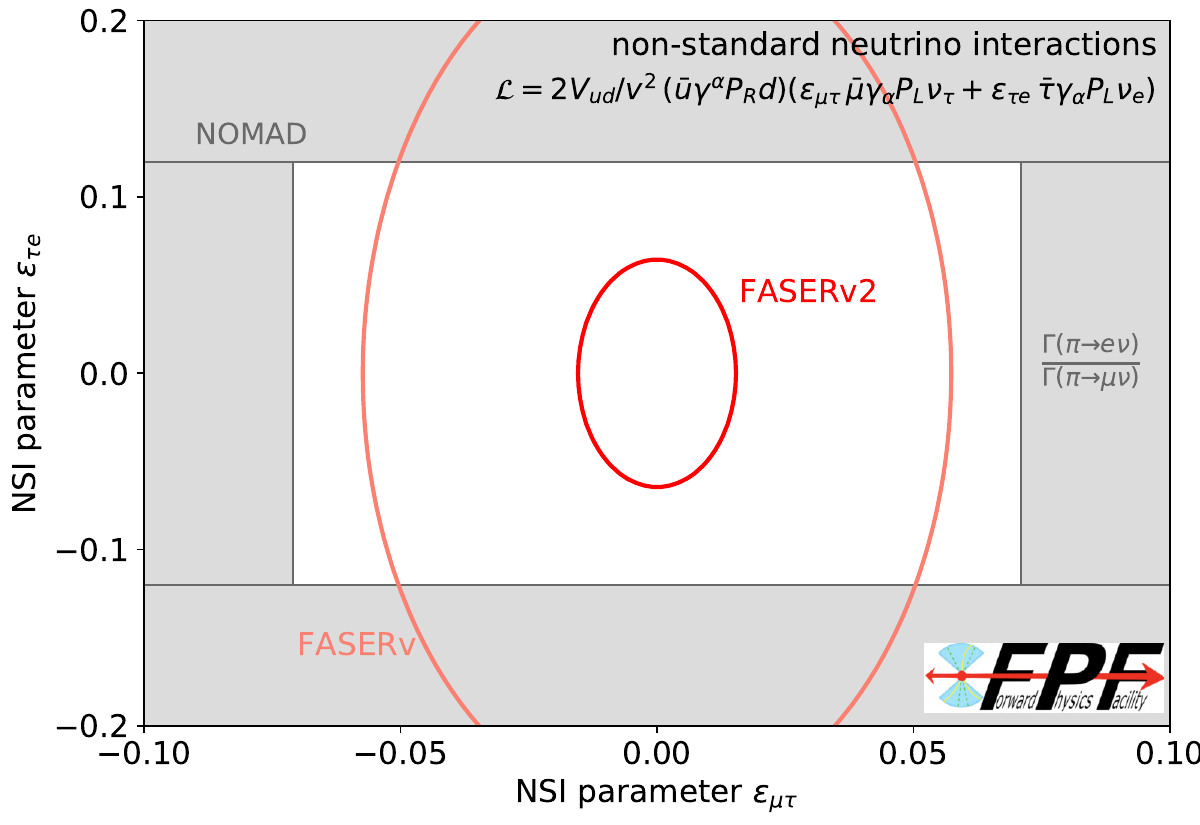}
  \vspace{-0.5cm}
  \caption{\textbf{Precision tau neutrino studies at the FPF.} The projected sensitivity of FASER$\nu$2 to neutrino NSI parameters that violate lepton flavor universality~\cite{Kling:2023tgr} Past NOMAD bounds~\cite{NOMAD:2001xxt,NOMAD:2003mqg} are presented based on Ref.~\cite{Biggio:2009nt}. Constraints from the measured ratio of pion decay widths to the electron and muon~\cite{ParticleDataGroup:2024cfk} are obtained based on Ref.~\cite{Falkowski:2021bkq}.}
  \label{fig:tau_neutrino}
  \vspace{-0.3cm}
\end{wrapfigure}

\noindent \textbf{Tau Neutrino Precision Measurements:} Although only a few handfuls of tau neutrino interactions have been identified by previous experiments, thousands of tau neutrinos will be interacting in the FPF detectors. The FASER$\nu$2 detector will be able to detect them, definitively observe the anti-tau neutrino for the first time, and open up a new window to an era of tau neutrino precision studies at TeV energies. In particular, these observations will enable tests of lepton flavor universality. Deviations from universality may be parameterized, for example, by neutrino non-standard interactions (NSI). The potential of LHC neutrino experiments to constrain charged current NSI using neutrino interaction measurements $\nu_\ell q \to \ell' q'$ has been studied in Ref.~\cite{Falkowski:2021bkq}. While many of the associated effective field theory operators are already well constrained by either precision meson decay measurement at flavor factories or LHC measurements probing the related processes $q q' \to \nu_\ell \ell'$, collider neutrino experiments have the potential to obtain world-leading constraints on operators associated to tau neutrinos by utilizing the sizable flux of tau neutrinos at the LHC. An example of FASER$\nu$2's sensitivity to probe two NSIs associated with tau neutrinos is shown in \cref{fig:tau_neutrino} as obtained in Ref.~\cite{Kling:2023tgr}. Here the $\epsilon_{\mu \tau}$ term leads to decay $\pi^+ \to \mu \nu_\tau$ while $\epsilon_{\tau e}$ induces the interaction $\nu_e d \to \tau u$. \medskip

\noindent \textbf{Neutrino-philic New Physics:} The large intensity and energy of the LHC neutrino beam at the LHC also provides a variety of novel opportunities to search for new physics. This includes searches for new neutrino-philic mediators that modify the predicted tau-neutrino flux at the FPF~\cite{Kling:2020iar, Batell:2021snh}; searches for modulinos~\cite{Anchordoqui:2021ghd, Anchordoqui:2023qxv} or sterile neutrinos with multi-eV masses~\cite{Anchordoqui:2024ynb}, leading to visible neutrino oscillation patterns for mass splittings $\Delta m^2 \sim 2500~\text{eV}^2$ way above those of short baseline or reactor neutrino experiments; searches for anomalous electromagnetic properties of neutrinos~\cite{MammenAbraham:2023psg}; searches for neutrino NSIs that modify neutrino production or neutrino scattering~\cite{Ismail:2020yqc, Falkowski:2021bkq, Kling:2023tgr}; searches for neutrino self-interactions and neutrino-philic mediators to DM~\cite{Kelly:2021mcd}; and constraints on BSM neutrino interactions through measurements of rare scattering processes, e.g., neutrino trident production~\cite{Altmannshofer:2024hqd}.

\subsection{QCD 
\label{sec:science_qcd}}

\begin{figure}[t]
\centering
\includegraphics[width=0.90\textwidth]{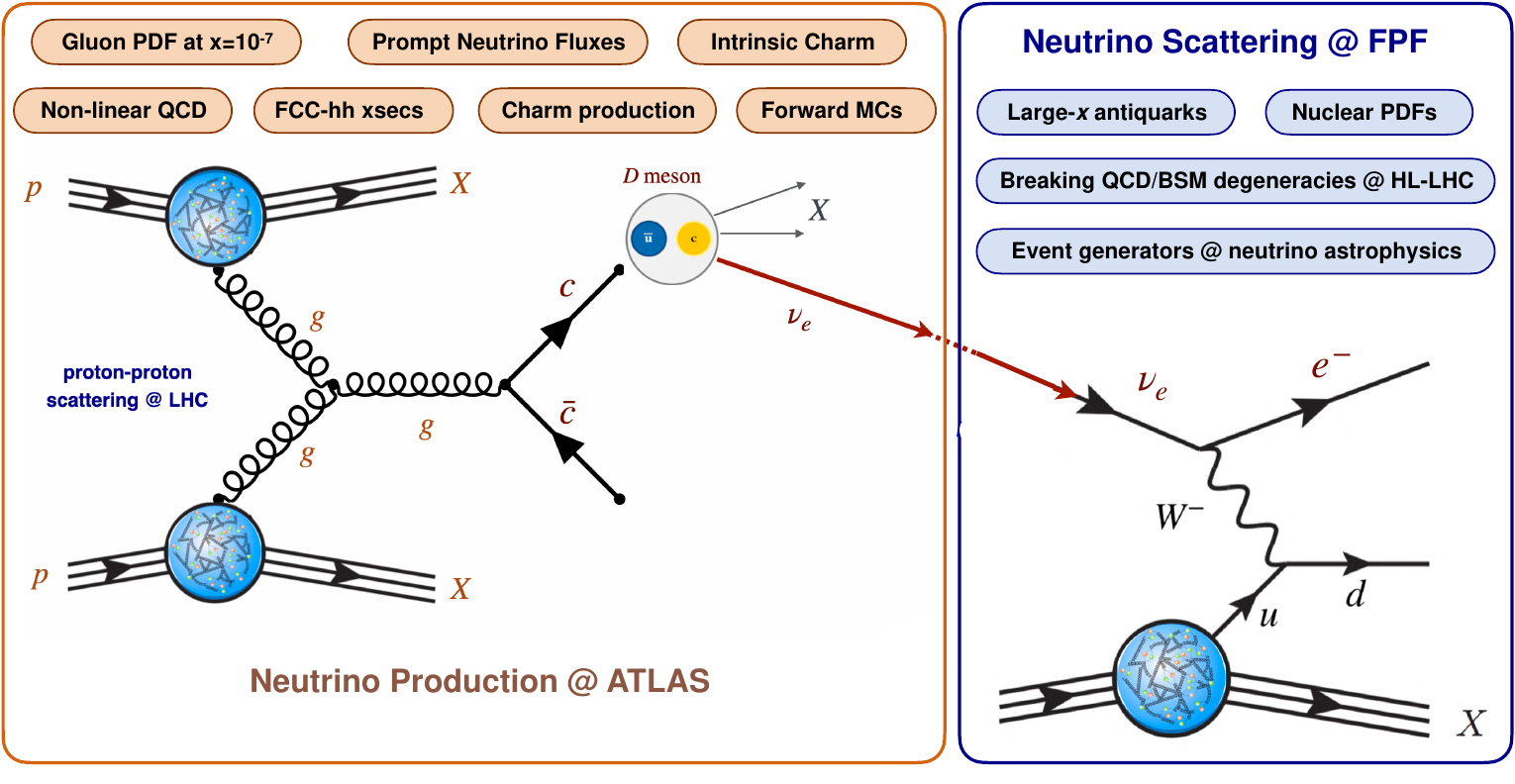}
\caption{\textbf{QCD physics at the FPF.} 
Representative QCD targets at the FPF, classified into production at the ATLAS IP and scattering at the FPF neutrino detectors.
}
\label{fig:FPF-SM-summary}
\end{figure}

The FPF offers unprecedented potential for innovative studies in QCD and hadronic structure.
Representative targets are summarised in Fig.~\ref{fig:FPF-SM-summary}, classified into whether sensitivity arises from  production at the ATLAS IP or from scattering at the FPF.
Neutrino production in $pp$ collisions constrains the gluon parton distribution function (PDF) down to $x\sim 10^{-7}$~\cite{Gauld:2016kpd,Rojo:2024tho}, charm production~\cite{Zenaiev:2019ktw}, forward hadron production~\cite{Kling:2023tgr}, non-linear QCD dynamics~\cite{Duwentaster:2023mbk,Bhattacharya:2023zei}, and intrinsic charm~\cite{Ball:2022qks,Maciula:2022lzk}, among other phenomena.
In turn, neutrino scattering at the FPF enable mapping large-$x$ nucleon structure~\cite{Cruz-Martinez:2023sdv}, breaking degeneracies between BSM signals and QCD effects in high-$p_T$ tails at the HL-LHC~\cite{Hammou:2023heg,Kassabov:2023hbm}, and tuning generators for neutrino astrophysics~\cite{vanBeekveld:2024ziz,Fieg:2023kld,FerrarioRavasio:2024kem,Buonocore:2024pdv}.
\medskip
 
\begin{figure}[b]
\centering
\includegraphics[width=0.99\textwidth]{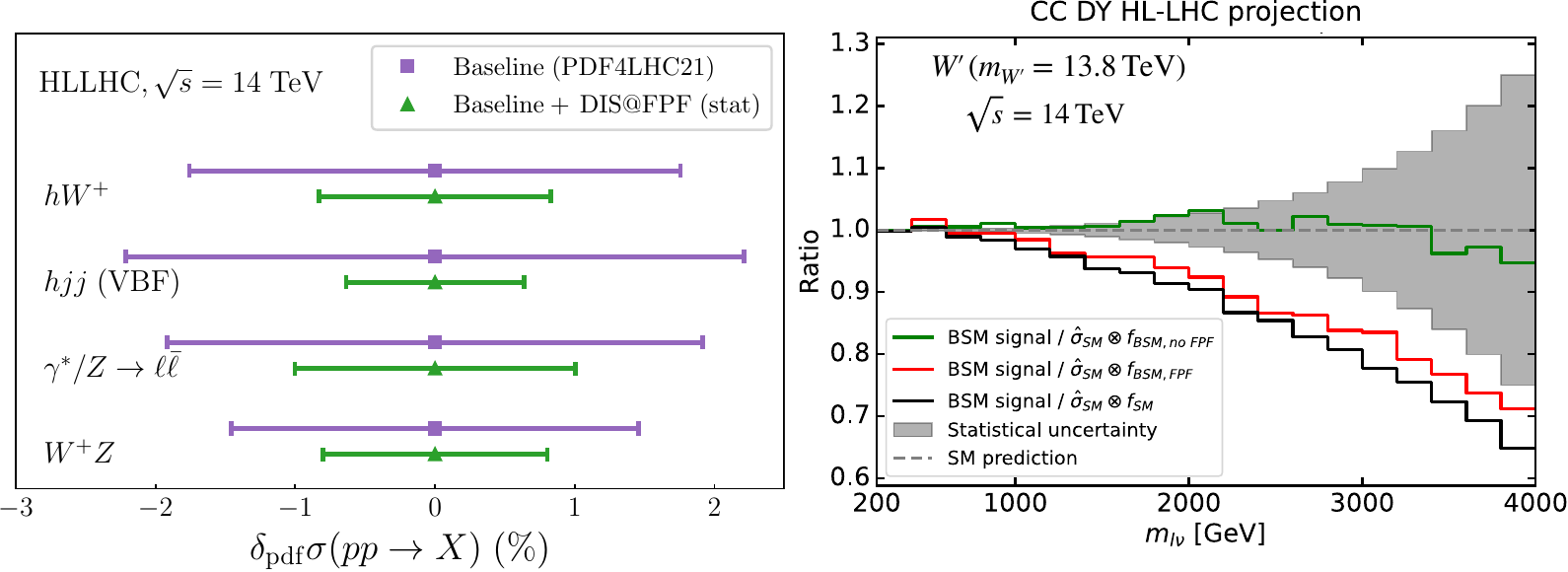}
\caption{
\textbf{Impact of FPF neutrino measurements on cross-section measurements and traditional BSM Searches at the HL-LHC.}
Left: Reduction of the PDF uncertainties on Higgs- and weak gauge-boson cross sections at the HL-LHC, enabled by neutrino DIS measurements at the FPF~\cite{Cruz-Martinez:2023sdv}.
Right: Signatures for a new heavy $W'$ boson with $m_{W'}=13.8$ TeV, namely a distortion of the high-mass charged-current Drell-Yan cross-sections, would be reabsorbed in a PDF fit with HL-LHC data ($f_{\rm BSM,noFPF}$), unless the PDFs are constrained by the ``low energy'' FPF neutrino data ($f_{\rm BSM,FPF}$)~\cite{Hammou:2024xuj}. 
}
\label{fig:FPF-HLLHC}
\end{figure}

\noindent
{\bf FPF Neutrino Measurements Enhance HL-LHC Discovery Prospects:}
Dedicated projections for neutrino DIS~\cite{Candido:2023utz} at the FPF~\cite{Cruz-Martinez:2023sdv} demonstrate that the expected $\mathcal{O}(10^5)$ electron-neutrino and $\mathcal{O}(10^6)$ muon-neutrino interactions provide stringent constraints on the proton PDFs.
DIS neutrino interactions are sensitive to momentum fractions $x \sim Q^2/(2 m_p E_\nu)$, where $Q^2$ is the momentum transfer and $m_p$ is the proton mass. 
This implies that the roughly ten times larger neutrino energies provided by the LHC allow to probe roughly ten times smaller values of the momentum fraction compared to neutrino scattering measurements at previous accelerator experiments such as NuTeV~\cite{NuTeV:2005wsg}.
Neutrino collisions at the FPF therefore cover a range in the $(x,Q^2)$ plane that overlaps with that of the Electron-Ion Collider (EIC)~\cite{AbdulKhalek:2021gbh}, while probing complementary flavor combinations.
For example, the muon charge identification and $D$-meson tagging capabilities in the FPF neutrino experiments allows the selection
of processes like $\nu s \to \mu^- c$ and $\bar\nu \bar s \to \mu^+ \bar c $ and therefore enable disentangling specific initial state quark and antiquark flavors including the relatively poorly constrained strange content of the proton. 
The sensitivity to improve PDFs via neutrino scattering measurements at the FPF was obtained in Ref.~\cite{Cruz-Martinez:2023sdv}.
The impact of DIS neutrino measurements at the FPF on the traditional HL-LHC program is twofold. 
On the one hand, FPF-constrained PDFs enable more precise theoretical predictions for core processes at the HL-LHC such as Higgs, Drell-Yan, and diboson production (left panel of Fig.~\ref{fig:FPF-HLLHC}). Measurements of these cross sections at ATLAS and CMS are therefore more sensitive to BSM physics.
On the other hand, if BSM signals are present in high-$p_T$ tails at the HL-LHC, they could inadvertently be reabsorbed into a PDF fit~\cite{Kassabov:2023hbm,Hammou:2023heg,Hammou:2024xuj}.

Breaking this degeneracy between QCD and BSM effects in high-energy scattering is possible by including the ``low-energy'' FPF data into the PDF fit.
For instance, assume a new heavy $W'$ boson with mass $m_{W'}=13.8$ TeV, outside the direct reach of the LHC.
The existence of this new  $W'$ boson would nevertheless distort the observable charged-current high-mass Drell-Yan cross-section in a kinematic region accessible at the HL-LHC.
As demonstrated in the right panel of Fig.~\ref{fig:FPF-HLLHC}, which shows the ratio $R$ between the injected BSM signal and different theory predictions based on the partonic SM cross-section, the effects of such $W'$ boson would be ``fitted away'' in a PDF fit including Drell-Yan data from the HL-LHC ($f_{\rm BSM,noFPF}$) unless the PDFs are constrained with the FPF neutrino data ($f_{\rm BSM,FPF}$)~\cite{Hammou:2023heg}. 
FPF data therefore greatly enhance the discovery potential of ATLAS and CMS for high mass particles. 
Neutrino DIS at the FPF also opens a novel window on the 40-year-old conundrum of whether intrinsic charm (or even bottom) quarks exist in the proton~\cite{Ball:2022qks,Guzzi:2022rca,Maciula:2022lzk}. \medskip

\begin{wrapfigure}{t}{0.49\textwidth}
  \centering
  \vspace{-0.5cm}
  \includegraphics[width=0.47\textwidth]{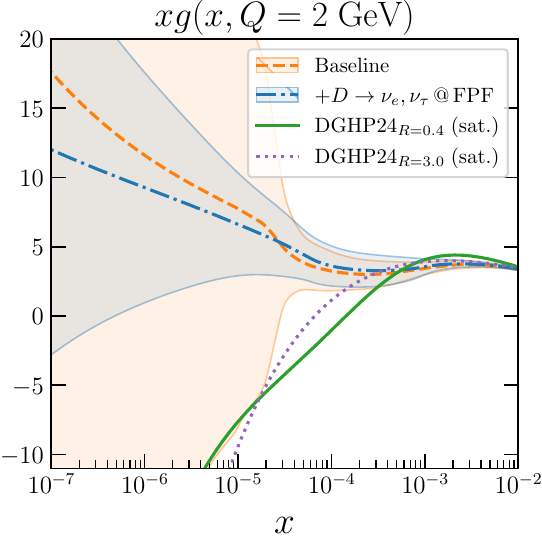}
  \vspace{-0.2cm}
  \caption{\textbf{Small-$x$ QCD at the FPF.} Impact of FPF data on the small-$x$ gluon PDF , compared with non-linear QCD (saturation) models. The $y$ axis displays $xg$ evaluated
  at a scale of $Q=2$ GeV. The baseline prediction is \texttt{NNPDF~3.1}~\cite{NNPDF:2017mvq}.}
  \label{fig:smallxQCD}
  \vspace{-0.2cm}
\end{wrapfigure}

\noindent
{\bf Small-$x$ QCD from Charm Production:}
The LHC neutrino fluxes depend sensitively on the mechanisms for forward light and heavy hadron production in $pp$ collisions~\cite{Kling:2021gos,Buonocore:2023kna}.
Both high energy electron neutrinos and tau neutrinos primarily originate from charm hadrons. 
These are mainly produced via gluon fusion, where one gluon carries a large momentum fraction $x\sim 1$ while the other carries a very small momentum fraction $x \sim 4 m_c / s \sim 10^{-7}$.
For comparison, measurements of forward $D$-meson production at LHCb can constrain the gluon PDF only down to $x\sim 10^{-5}$~\cite{Gauld:2016kpd,Zenaiev:2019ktw}.
By defining tailored observables where theory uncertainties cancel out, such as the ratio between electron and tau neutrino event rates, FPF measurements can be used to pin down the gluon PDF down to $x\sim 10^{-7}$~\cite{Rojo:2024tho}, as shown in Fig.~\ref{fig:smallxQCD}.
Such measurements inform the study of novel QCD dynamics at small-$x$, a region where non-linear and BFKL-like effects are expected to dominate, as highlighted by the DGHP24 predictions~\cite{Duwentaster:2023mbk} for the gluon PDF based on saturation (recombination) effects built into the DGLAP evolution.
Constraints on the small-$x$ gluon PDF would be instrumental to inform FCC-hh cross sections, since at $\sqrt{s}=100$ TeV even Higgs and gauge boson production becomes a ``small-$x$'' process with potentially large corrections from BFKL resummation~\cite{Bonvini:2018vzv,Rojo:2016kwu}.
These constraints on small-$x$ QCD are also relevant for astroparticle physics, as further discussed below.
We emphasize that only in LHC neutrino experiments can one access this small-$x$ region crucial to probe QCD and astroparticle physics processes, e.g. beam dump experiments such as SHiP~\cite{SHiP:2015vad} involve neutrinos of much lower energy (tens of GeV) which hence can access only a region $x> 10^{-2}$ both in production and in scattering.
Other neutrino experiments, such as DUNE, involve yet smaller energies, where the DIS component is essentially negligible. 
\medskip

\noindent {\bf Neutrino Event Generators:}
The robust interpretation of FPF measurements demands state-of-the-art Monte Carlo event generators for neutrino scattering at TeV energies.
Such generators, based on higher-order QCD corrections and matched to modern parton showers, are also relevant to model high-energy neutrino scattering at neutrino telescopes such as IceCube~\cite{IceCube:2016zyt} and KM3NeT~\cite{Coniglione:2015aqa}. 
Testing and validating  neutrino event generators, such as the {\sc\small POWHEG}-based ones presented in Refs.~\cite{vanBeekveld:2024ziz,FerrarioRavasio:2024kem,Buonocore:2024pdv}, on FPF data is also instrumental  for the FPF BSM program, with neutrino signals representing the leading background in many searches.
Measurements of fragmentation functions in neutrino DIS also probe the cold nuclear medium of the target nucleus, complementing $eA$ scattering analyses at the EIC. 

\subsection{Astroparticle Physics 
\label{sec:science_app}}

Besides addressing key questions in astrophysics, high-energy cosmic-ray and neutrino experiments provide unique access to particle physics at center-of-mass energies that are an order of magnitude higher than LHC $pp$ collisions~\cite{Anchordoqui:2018qom,Albrecht:2021cxw}.
The FPF provides unique opportunities for interdisciplinary studies at the intersection of  particle and astroparticle physics~\cite{Feng:2022inv,Soldin:2023gox,Soldin:2024iev,Anchordoqui:2022ivb}.
\medskip

\begin{figure}[tbh]
\includegraphics[angle=0,width=0.99\textwidth]{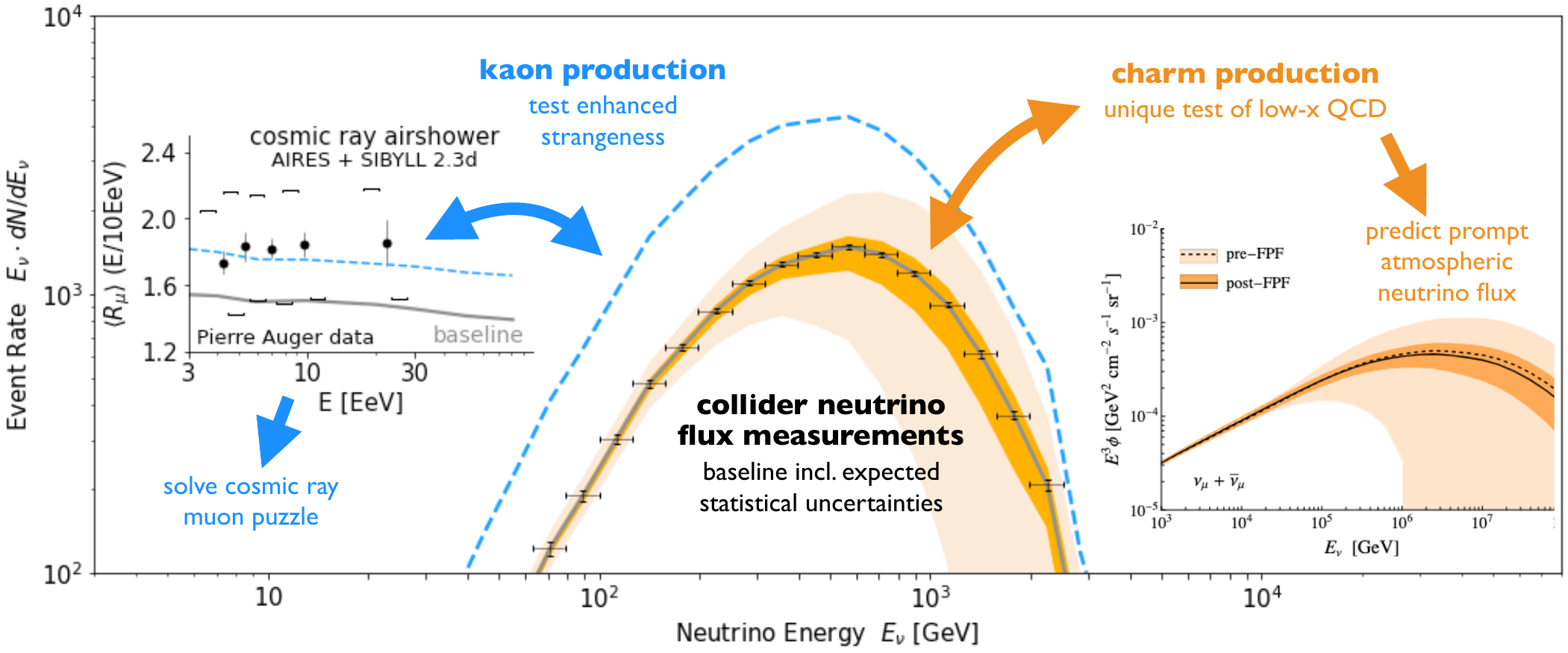}
\caption{\textbf{Astroparticle physics at the FPF.} 
The central part of the figure shows the expected energy spectrum of interacting electron neutrinos in the FLArE detector at the FPF (solid gray curve) obtained using \texttt{SIBYLL~2.3d}~\cite{Riehn:2019jet} and \texttt{POWHEG + Pythia}~\cite{Buonocore:2023kna} with the \texttt{NNPDF~3.1} as well as expected statistical uncertainties (black error bars).
The colored contours illustrate two examples of physics that can change the expected flux and be probed at the FPF: enhanced kaon production that solves the muon puzzle (blue dashed line) and small-x PDFs that lead to improved prompt atmospheric neutrino flux predictions (orange band). 
Left: Dimensionless muon shower content $R_\mu$ as predicted by {\tt piKswap} model through simulations with {\tt SIBYLL 2.3 + AIRES}~\cite{Sciutto:1999jh} and compared with data from the Pierre Auger Observatory~\cite{PierreAuger:2014ucz}; for details, see~\cite{Anchordoqui:2022fpn}. 
Right: Reduction of PDF uncertainties on the prompt neutrino flux $\Phi$ enabled by FPF data as a function of $E_\nu$; see \cref{fig:smallxQCD} for the corresponding PDF.}
\label{fig:APP-summary}
\end{figure}

\noindent
{\bf The Muon Puzzle:}
For many years, the experimental measurements of the number of muons in high- and ultra-high-energy cosmic-ray air showers have appeared to be in tension with model predictions~\cite{PierreAuger:2014ucz, PierreAuger:2016nfk,TelescopeArray:2018eph,EAS-MSU:2019kmv, Soldin:2021wyv,PierreAuger:2024neu}. This conundrum has been dubbed the cosmic-ray muon puzzle. Various air-shower models~\cite{Riehn:2019jet,Fedynitch:2018cbl,Riehn:2024prp,Pierog:2013ria,Ostapchenko:2013pia,Ostapchenko:2005nj,Roesler:2000he} can be tested under controlled experimental conditions at the FPF, because the ratio of the low-energy electron and muon neutrino fluxes is a proxy for the charged kaon to pion production rate. The differences in the predicted fluxes exceed a factor of two, which is much larger than the expected statistical uncertainties at the FPF~\cite{Kling:2021gos}. Since the muon puzzle is assumed to be of soft-QCD origin~\cite{Albrecht:2021cxw}, there is also a strong connection to the QCD program of the FPF and dedicated QCD measurements will further help to understand particle production in cosmic-ray air showers. Thorough analyses have suggested that an enhanced rate of strangeness production in the forward direction could explain the observed discrepancies~\cite{Allen:2013hfa, Anchordoqui:2016oxy,Anchordoqui:2019laz,Albrecht:2021cxw}. A specific example accounting for enhanced strangeness production that can resolve the muon puzzle is the simple phenomenological (one-parameter) {\tt piKswap} model~\cite{Anchordoqui:2022fpn, AbdulKhalek:2018rok}. It predicts a significant increase of electron neutrinos with energies below 1~TeV that can be tested at the FPF, as illustrated by the blue curve in \cref{fig:APP-summary}.
\medskip

\noindent
{\bf Atmospheric Neutrino Fluxes:}
High-energy neutrinos of astrophysical origin are routinely observed by large-scale neutrino telescopes, such as IceCube~\cite{IceCube:2016zyt} and KM3NeT~\cite{Coniglione:2015aqa}, and atmospheric neutrinos produced in extensive air showers are an irreducible background for these searches. Neutrinos at high energies above 1~TeV are mainly produced in charm hadron decays. 
If the charm PDF in the proton is entirely perturbative, the production of $D$-mesons in cosmic ray collisions is dominated by gluon fusion ($gg\to c\bar{c}$) and can be described using perturbative QCD~\cite{Gauld:2015kvh}.
In the presence of a significant intrinsic charm PDF component~\cite{Ball:2022qks} instead, the partonic reaction $cg\to cg$ dominates the calculation at high energies~\cite{Maciula:2022lzk}.
Measurements of the neutrino flux at the FPF therefore provide access to both the very high-$x$ and the very low-$x$ regions of the colliding protons. These measurements yield information about high-$x$ PDFs, in particular intrinsic charm, as well as novel QCD production mechanisms, such as BFKL effects and non-linear dynamics, well beyond the coverage of other experiments and providing key inputs for astroparticle physics. 
FPF measurements will constrain the underlying PDFs and therefore provide stringent constraints on the prompt atmospheric neutrino flux, contributing to the scientific program of large-scale neutrino telescopes.
This is further quantified in \cref{fig:APP-summary}, showing theoretical predictions for the prompt muon-neutrino flux based on the formalism of Refs.~\cite{Bai:2022xad,Bai:2021ira}, considering only PDF uncertainties, before and after FPF constraints are included. 
Although other sources of theory uncertainty contribute to the total error budget, \cref{fig:APP-summary} demonstrates the strong sensitivity of the FPF to the mechanisms governing atmospheric neutrino production from charm decays.

\section{The Facility 
\label{sec:wg0}}

The FPF facility has been studied by CERN experts over the last four years, with technical studies detailed in Refs.~\cite{PBCnote,PBCnote2,vibration-note}. 
The work has benefited from the vast experience at CERN in designing and implementing many similar large underground facilities, particularly the recent HL-LHC underground works at the ATLAS and CMS IPs. Many of the same technical solutions can be adopted for the FPF, and lessons learned can also be applied. \medskip

\begin{figure}[t]
\centering
\includegraphics[width=0.68\textwidth, height=0.35\textwidth]{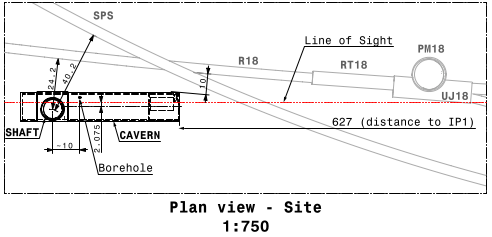}
\hfill
\includegraphics[width=0.27\textwidth]{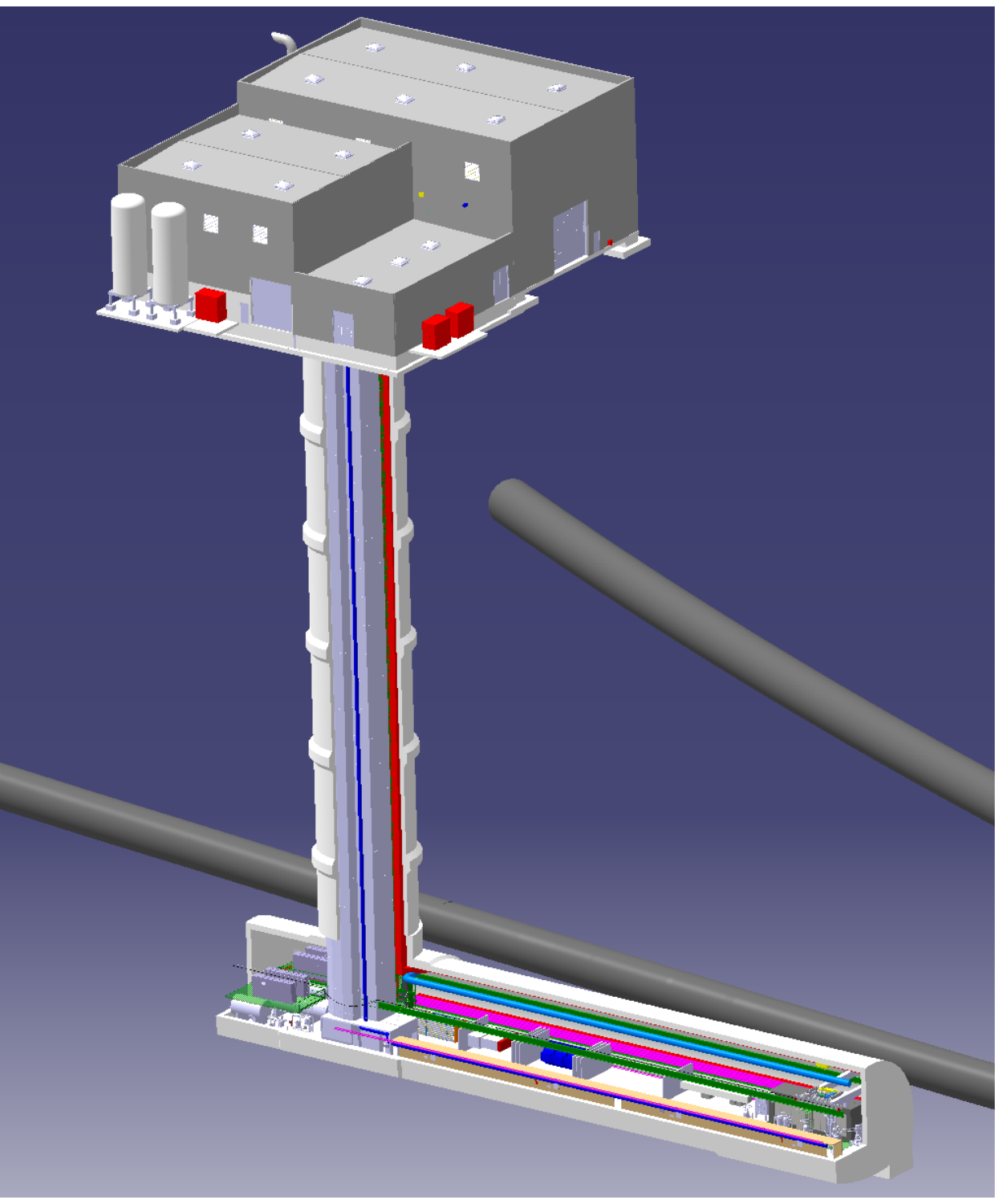}
\caption{Left: Plan view showing the FPF location.  Right: 3D view of the Facility. All distances are given in meters.
}
\label{fig:plan}
\end{figure}

\begin{figure}[b]
\centering
\includegraphics[width=0.98\textwidth]{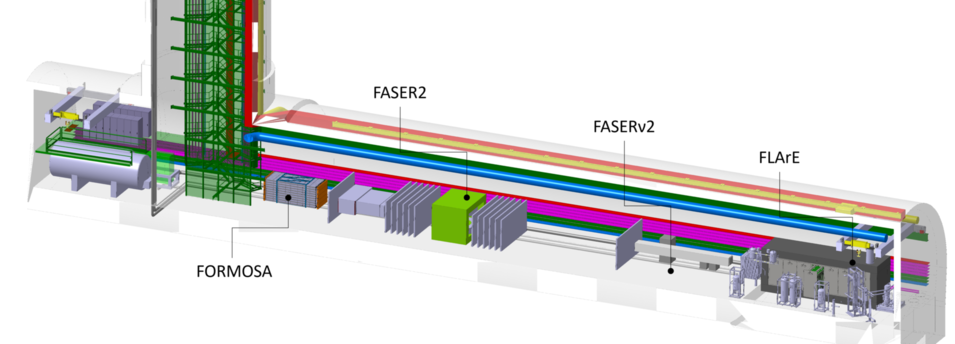}
\caption{The baseline layout of the FPF facility, showing the four proposed experiments and the large infrastructure. 
}
\label{fig:facility-layout}
\end{figure}

\noindent \textbf{Site Selection and Cavern Design:} A site optimization to find the best location for the FPF facility was carried out. This identified an optimal site 627~m west of the ATLAS IP (IP1), on CERN land in France, as shown in \cref{fig:plan}. Following this, the facility design has been through several iterations to optimize the layout for the proposed detectors, along with the needed technical infrastructure. The current baseline design is shown in \cref{fig:facility-layout}. This includes a 75~m-long, 12~m-wide underground cavern, with a dedicated experimental area (65~m long) and a service cavern (5~m long), as well as an 88~m-deep shaft and the associated surface building for access and services. The closest point between the underground cavern and the LHC tunnel is 10~m, as required by the civil engineering and radiation protection teams. \medskip

\noindent \textbf{Site Investigation and Geological Conditions:} In Spring 2023, a site investigation study was carried out where a 20~cm-diameter, 100~m-deep core was drilled at the proposed location of the FPF shaft. Analysis of the extracted core confirmed that the geology is good for the planned excavation works, and no show stoppers were identified. 
A Class 4 costing for the civil engineering work has been carried out, based on similar work carried out at CERN in the last decade and taking into account the findings of the site investigation. 
The costing methodology has been cross checked by an external civil engineering consultant. The cost estimate is 35~MCHF for the underground works, shaft, and surface buildings.  A detailed breakdown of these costs is given in \cref{sec:appendix}. The expected time for the civil engineering works is 3~years.\medskip

\noindent \textbf{Excavation Work and Vibrations:} The possibility of carrying out the FPF excavation work during beam operation will allow much more flexibility in the FPF implementation schedule. However, concerns have been raised that the excavation works could impact beam operations of the LHC or SPS, leading to beam losses and possible beam dumps. The CERN accelerator group has carried out detailed studies of the effect of the expected vibration level from the excavation on beam operation performance, as documented in Ref.~\cite{vibration-note}. The conclusion of these studies is that no problems are foreseen, and the excavation can be carried out during beam operations. \medskip 

\begin{figure}[th]
\centering
\includegraphics[trim=0.25cm 0cm 0.25cm 2.cm,clip,width=0.49\textwidth]{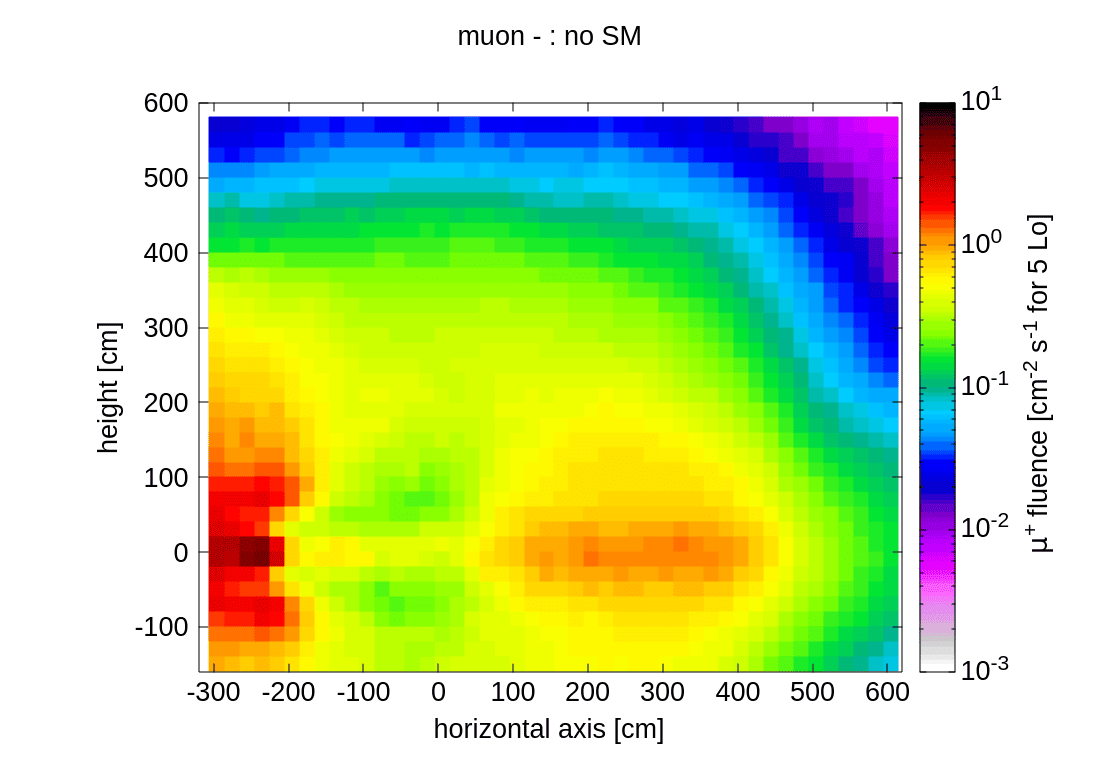}
\includegraphics[trim=0.25cm 0cm 0.25cm 2.cm,clip,width=0.49\textwidth]{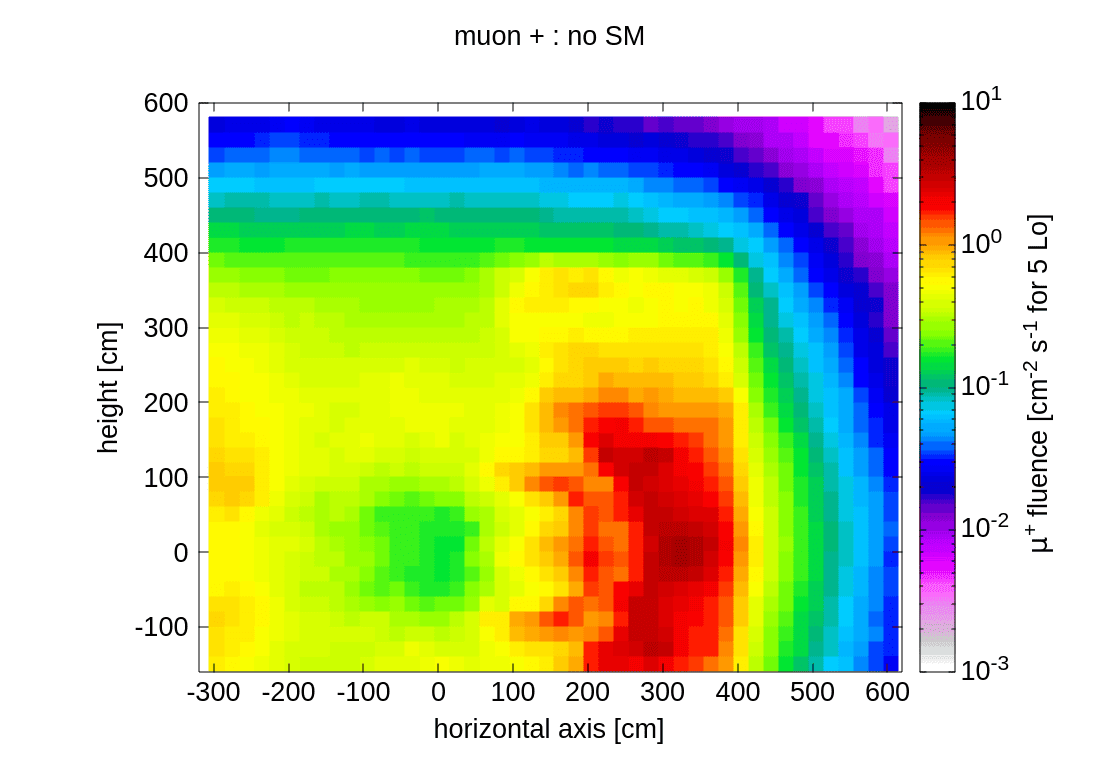}
\caption{The muon fluence rate for $\mu^-$ (left) and $\mu^+$ (right) in the transverse plane in the FPF cavern for the HL-LHC baseline luminosity of $5 \times 10^{34}$~cm$^{-2}$~s$^{-1}$. The coordinate system is defined such that $(0,0)$ is the LOS, and -ve x is towards the center of the LHC ring. }
\label{fig:mu-flux}
\end{figure}

\noindent \textbf{Muon Fluxes:} The expected muon background rate in the FPF has been estimated using FLUKA~\cite{fluka} simulations. These simulations include a detailed description of the infrastructure between IP1 and the FPF. For the LHC Run 3 setup, the simulations have been validated at the $\mathcal{O}$(25\%) level with FASER~\cite{FASER:2018bac} and SND@LHC~\cite{SNDLHC:2023mib} data.  However, for the HL-LHC, much of the accelerator infrastructure (magnets, absorbers, etc.) in the relevant region will change. As shown in \cref{fig:mu-flux}, for the baseline HL-LHC luminosity of $5 \times 10^{34}$ cm$^{-2}$ s$^{-1}$, FLUKA simulations predict a muon flux of 0.6~cm$^{-2}$~s$^{-1}$ within 50~cm of the LOS, with the flux substantially higher when going to 2~m from the LOS in the horizontal plane, as can be seen in the figure. In general, the expected muon rate is acceptable for the proposed experiments, however reducing the rate would be beneficial. Studies on the effectiveness of installing a sweeper magnet in the LHC tunnel or using the beam corrector magnets to reduce the flux are ongoing.  \medskip

\noindent \textbf{Radiation Levels and Safety:} FLUKA simulations have also been used to assess the radiation level relevant for the detectors, which is estimated assuming the HL-LHC baseline integrated luminosity of 360~fb$^{-1}$ per year. 
The neutron field that can cause radiation-induced damage to silicon detectors has been assessed to be less than 
10$^7 n_{\rm eq}$ / cm$^{2}$/y (where $n_{\rm eq}$ is the  Silicon 1 MeV neutron equivalent fluence). This is many orders of magnitude lower than that in the LHC experiments.   
The annual high-energy hadron equivalent fluence that determines the single event error rate in electronics, does not exceed $3 \times 10^6$~cm$^{-2}$~y$^{-1}$, which is the threshold adopted in the LHC for declaring an area safe from the radiation to electronics (R2E) point of view~\cite{R2E}.

Being able to access the cavern during beam operations will be extremely valuable for detector installation, commissioning, and maintenance tasks. It will also allow the experiments to be upgraded or even replaced, as may be necessary to respond to the evolution of the physics landscape over the time period of the HL-LHC. FLUKA simulations have been used to assess the radiation level in the FPF cavern during beam operation. These studies show that the radiation source will be solely from muon-induced particles. The expected radiation level will be low enough for people to access the cavern during beam operation, provided they are trained as radiation workers, carry a dosimeter, and are there for less than 20\% of the time integrated over a year. However, some parts of the cavern may be classified as local short stay areas.  \medskip

\noindent \textbf{Transport and Detector Integration:} Integration studies have shown that the proposed experiments (in their current form) can be installed and fit into the baseline cavern, including their main associated infrastructure. 
Standard infrastructure and services that have been considered so far include cranes and handling infrastructure, electrical power, ventilation systems, fire/smoke safety, access, and evacuation systems. A very preliminary costing of these services (based on existing CERN standard solutions) is at the level of less than 10~MCHF, giving a total costing of the facility, including both civil engineering and outfitting, of around 45~MCHF.

\section{FASER2 
\label{sec:wg5}}

FASER2 is a large-volume detector comprised of a spectrometer, electromagnetic and hadronic calorimeters, veto detectors and a muon detector, that is designed for sensitivity to a wide variety of models of BSM physics and for precise electron and muon reconstruction for neutrino measurements. It builds on positive experience gained from the successful operation of the existing FASER experiment~\cite{FASER:2022hcn}, a much smaller detector, which was constrained to be situated within an LHC transfer tunnel. The FASER2 detector, specifically designed for the FPF facility, is much larger (by a factor of $\sim 600$ in decay volume size) and includes new detector elements. It provides an increase in reach for various BSM signals of several orders of magnitude compared to FASER and allows sensitivity to models that were previously out of reach, such as dark Higgs, heavy neutral lepton, and axion-like particle models, as studied in Refs.~\cite{FASER:2018ceo,FASER:2020gpr,FASER:2018eoc}. There is unique sensitivity in inelastic dark matter and in searches for quirks in a mass range motivated by naturalness arguments, as discussed in \cref{sec:physics}.

In addition to the BSM case for FASER2, the SM neutrino program at the FPF will rely on the identification of muons from neutrino decays and precise measurement of their momentum and charge. The FASER2 spectrometer will be integral for these measurements for both FASER$\nu$2 and FLArE. 

\begin{figure}[tb]
  \centering
  \includegraphics[width=0.99\textwidth]{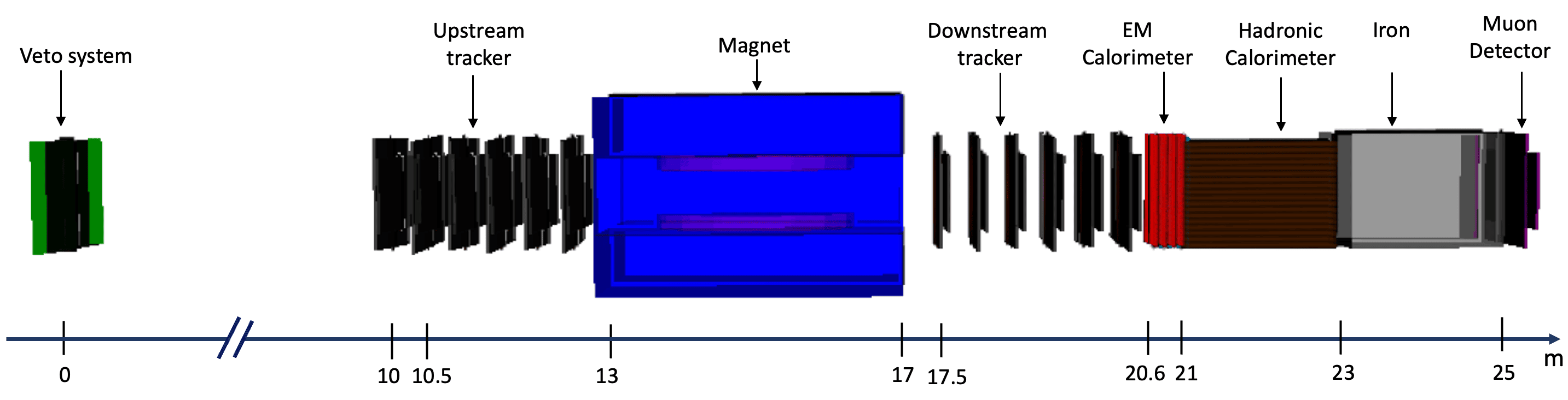}  
  \caption{Visualisation of the full FASER2 detector, showing the veto system, uninstrumented 10~m decay volume, tracker, magnet, electromagnetic calorimeter,  hadronic calorimeter, iron absorber and muon detector. 
  \label{fig:FASER2-Design}}
\end{figure}

Figure~\ref{fig:FASER2-Design} shows a rendering of the GEANT4 model of the full FASER2 detector. This design is the result of several iterations and improvements, but it is still a work in progress. The overall layout is largely driven by the spectrometer, which is itself constrained by considerations relating to deliverable and affordable magnet technology. This leads to a baseline detector configuration consisting of a spectrometer with a large-volume dipole magnet. The magnet has a rectangular aperture of 1~m in height and 3~m in width. This also defines the transverse size of the decay volume, which is the 10~m uninstrumented region upstream of the first tracking station (a $2.6 \times 1 \times 10$~m$^3$ cuboid) and downstream of the veto station. Maximising the transverse size is a general design requirement driven by the need to have sufficient acceptance for BSM particles originating from heavy flavor decays and charged leptons arising from neutrino interactions in FLArE. Studies are ongoing as to whether the decay volume would need to be under vacuum or filled with low-density gas, i.e., helium, to achieve background-free measurements. A more square (e.g., $1.7 \times 1.7$~m$^2$) aperture is also under consideration, as it improves the acceptance of muons from FLArE by 5-10\% without a significant degradation in LLP sensitivity. \medskip

The baseline integrated magnetic field strength is 2~Tm. This is optimised based on simulations that demonstrate that the required charged particle separation, momentum resolution, and charge identification are obtained for the BSM and neutrino programme, while keeping the field strength to an acceptable minimum to reduce cost. Superconducting magnet technology is required to maintain such a field strength across a large aperture. Recent investigations by KEK magnet experts, along with discussions with manufacturing experts at Toshiba in Japan and Tesla Engineering in the UK, have demonstrated that this design is feasible at an acceptable cost ($\sim 4$ MCHF) and lead-time (3.5 years).  Alternative options are also being investigated to make use of industrial magnets with a smaller aperture (circular with 1.6 m diameter) and lower field strength ($\sim 1.5$ Tm). These magnets are commercially available at a lower cost, and while they do lead to a limited degradation in sensitivity, this is not significant enough to put the main physics goals out of reach.  \medskip

For most FASER2 sub-detectors, a performant baseline is achievable from simpler well-understood detector technologies that will allow the major physics goals to be achieved. However, more advanced technologies are also under consideration to augment these baseline capabilities, and the evaluation of these has undergone the most scrutiny so far due to the higher associated cost. Such augmentations are especially appealing in the case that they can come via existing R\&D activities, for example, in the context of future colliders, where FASER2 can act as a mid-term testbed.

The tracking detectors are foreseen to use a SiPM and scintillating fiber tracker technology, based on LHCb's SciFi detector~\cite{Hopchev:2017tee}. This technology gives sufficient spatial resolution ($\sim 100\;\mu$m) at a significantly reduced cost compared to silicon detectors. The use of silicon-based tracking detectors will be explored for the interface between FASER2 and FASER$\nu$2, and for the first tracking station downstream of the decay volume. Possible augmentation utilising the LHCb MightyPix technology~\cite{Hennessy:2024jov} is under investigation for potential improvement in particle separation power in both the first tracker layer and in the central region of the transverse plane, where the LLP energy is higher and decay products more collimated. 

A simple lead-scintillator calorimeter would be sufficient for the reconstruction of energy deposits from electrons and hadronic decay products of LLPs. A more advanced calorimeter is also under study to be based on dual-readout calorimetry~\cite{Lee:2017xss,Antonello:2018sna} technology, especially for the central region. This builds upon experience of existing prototypes for future collider R\&D, but modified for the specific physics needs of FASER2: spatial resolution sufficient to identify particles at $\sim 1-10$~mm separation; good energy resolution; improved longitudinal segmentation with respect to FASER; and the capability to perform particle identification, separating, for example, electrons and pions. 

The ability to separately identify electrons and muons would be very important for signal characterization, background suppression, and for the interface with FASER$\nu$2. To achieve this, ${\cal O}(10)$ interaction lengths of iron will be placed after the calorimeter, with sufficient depth to absorb pions and other hadrons, followed by a detector for muon identification, for which additional SciFi planes could be used. 
Finally, the front veto system will be required to reject a rate of approximately 20 kHz of muons from the IP. Scintillator-based approaches have proven to be sufficient for this in FASER, and a similar, but re-optimised, design is foreseen for FASER2. 
The event rate and size are much lower than most LHC experiments, so the trigger needs are not expected to be a limiting issue. For instance, it is expected that it will be possible to significantly simplify the readout of the tracker, with respect to what is used in the LHCb SciFi detector.

Various performance studies have been performed to assess different design considerations and technologies for FASER2. Metrics such as momentum resolution, LLP sensitivity, and geometrical acceptance have been studied both in terms of physics performance and the implied detector technology complexity and cost. Different simulation tools have been utilised for these studies:~the FORESEE~\cite{Kling:2021fwx} package is used for the simulation and event generation of LLP production from forward hadrons; the Geant4~\cite{Agostinelli:2002hh} simulation framework is used for the propagation of particles through a magnetic field in the LLP decay product separation studied; and the ACTS~\cite{Ai:2021ghi} tool is used for track reconstruction studies.

An illustration of such studies is provided in the following for the expected momentum resolution. For the baseline detector outlined above, with an intrinsic resolution of 100~$\mu$m and 2 Tm integrated field strength, a muon momentum resolution of approximately 2(4)\% is achieved for 1(5)~TeV muons. This is expected to be sufficient for the physics goals of FASER2. Studies show the baseline design to be quite robust:~this performance is stable under a range of magnetic field strengths, and appreciable degradation only appears with a significantly worse intrinsic resolution. The momentum resolution was also studied as a function of the amount of material in each tracker layer and only when approaching an interaction length is a significant loss in resolution observed. Studies were also performed to understand the possible impact of detector misalignnment. This shows that significant misalignments can be corrected using a track-based alignment method, with a precision of $\sim 50$ $\mu$m obtainable. 

The FASER2 experiment will be essential to maximise the physics potential of the FPF. The baseline detector design has been optimised to obtain the required physics performance in an affordable way, but several systems could be upgraded to improve the performance at higher cost. Given the importance of the FASER2 magnet in the design, significant work has been carried out to find a baseline solution for this, with an alternative option using commercially-available magnet units also being considered.

\section{FASER$\nu$2 
\label{sec:wg6}}

FASER$\nu$2 is a 20-ton neutrino detector located on the LOS, a much larger successor to the FASER$\nu$~\cite{FASER:2019dxq} detector in the FASER experiment. With the FASER$\nu$ detector, the first evidence for neutrino interaction candidates produced at the LHC was reported in 2021~\cite{FASER:2021mtu}, and the first measurements of the $\nu_e$ and $\nu_\mu$ interaction cross sections at TeV energies were reported in 2024~\cite{FASER:2024hoe}. These results confirms the FASER$\nu$ emulsion detector's ability to deliver physics measurements in the LHC environment. 

An emulsion-based detector will identify heavy flavor particles produced in neutrino interactions, including tau leptons and charm and beauty particles. 
FASER$\nu$2 can perform precision tau neutrino measurements and heavy flavor physics studies, testing lepton universality in neutrino scattering and new physics effects, as well as providing important input to QCD and astroparticle physics, as described in \cref{sec:physics}. 

\begin{figure}[b]
\centering
\includegraphics[width=0.48\textwidth]{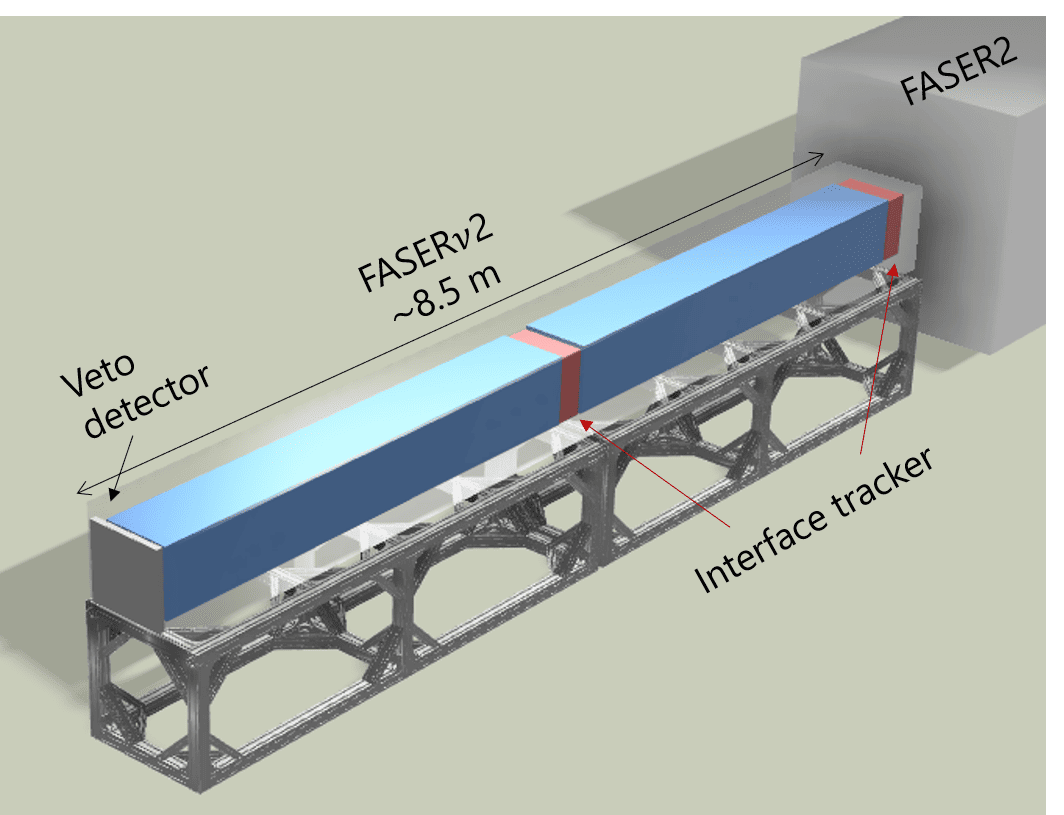}
\includegraphics[width=0.48\textwidth]{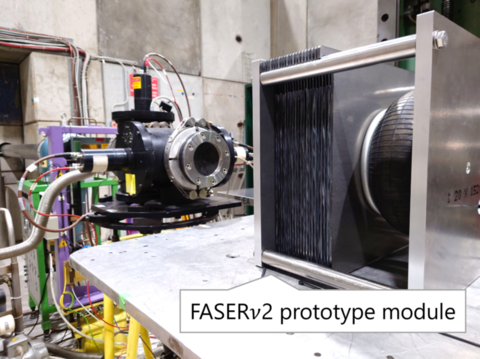}
\caption{Left: Design of the FASER$\nu$2 detector. Right: FASER$\nu$2 prototype module on the SPS-H8 beamline.}
\label{fig:fasernu2_schematics}
\end{figure}

The left panel of \cref{fig:fasernu2_schematics} shows a schematic of the proposed FASER$\nu$2 detector, which is composed of 3300 emulsion layers interleaved with 2-mm-thick tungsten plates. The total volume of the tungsten target is 40 cm $\times$ 40 cm $\times$ 6.6 m, with a mass of 20 tons. 
The emulsion detectors will be placed in cooling boxes and kept at around $10^\circ$C to avoid fading of the recorded signal. 
The detector will be placed directly in front of the FASER2 spectrometer along the LOS. The FASER$\nu$2 detector will also include a veto system and interface detectors to the FASER2 spectrometer, with one interface detector in the middle of the emulsion modules and the other detector downstream of the emulsion modules. These additional systems will enable a FASER2-FASER$\nu$2 global analysis and make measurement of the muon charge possible, a prerequisite for $\nu_\tau$/$\bar\nu_\tau$ separation. 
The veto system will be scintillator-based, and the interface detectors could be based on silicon strip sensors or scintillating fiber tracker technology. The detector length, including the emulsion films and interface detectors, will be approximately 8.5 m. 

A mechanical prototype has been produced to test critical technical challenges, namely applying pressure to fix sub-micrometer alignment and assembly under room light in the FPF experimental hall. As shown in the right panel of \cref{fig:fasernu2_schematics}, a test beam experiment was performed in July 2024 at the SPS-H8 beamline, confirming the concept of the techniques. 
In addition to the test of a mechanical prototype, other test samples were produced and exposed to the beam. With these, one can check the long-term performance of emulsion films to take data through a year without replacing emulsion films. One can also test a new type of photo-development solution, which increases the gain of chemical amplification, which can help maximize the readout speed of the emulsion detector. The analysis of the samples is ongoing.

\begin{figure}[tb]
\centering
\includegraphics[width=\textwidth]{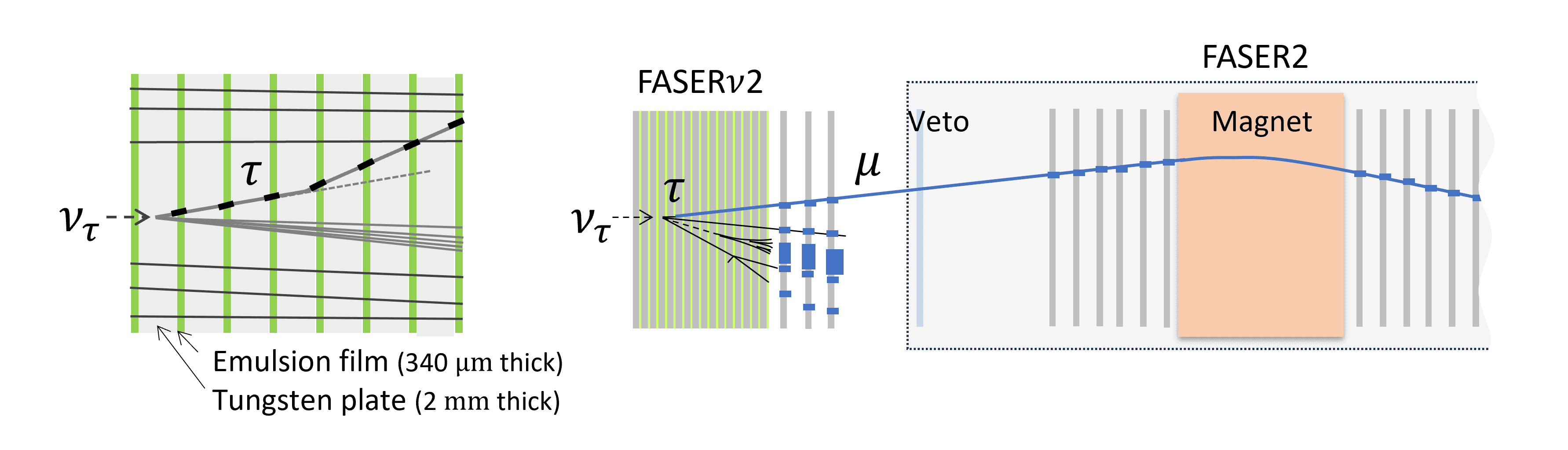}
\caption{Left: Tau decay topology in the emulsion detector. Right: Charge measurement for a muon from a tau decay.}
\label{fig:tau_to_mu}
\end{figure}

As the $c\tau$ of the tau lepton is 87 $\mu$m, a high-precision emulsion detector~\cite{Ariga:2020lbq} is essential to detect tau decays topologically. After optimizations of the detector performance in terms of precision, sensitivity, and long-term stability, emulsion gel with silver bromide crystals of 200~nm diameter will be used, which provides an intrinsic position resolution of 50~nm.  The left panel of \cref{fig:tau_to_mu} shows a tau decay topology in the emulsion detector. As shown in the right panel of \cref{fig:tau_to_mu}, a global analysis that links information from FASER$\nu$2 with the FASER2 spectrometer enables charge measurements of muons from tau decays, and thereby the detection of $\bar\nu_\tau$ for the first time.

Emulsion detector analysis will be limited by the accumulated track density and become difficult above $10^6$ tracks/cm$^2$ with the current tracking algorithms. 
To address the high track density in emulsion trackers caused by muon backgrounds, R\&D efforts focus on hardware, image processing, and reconstruction. At the hardware level, optimizing silver bromide crystal size and revising photo-development chemicals aim to reduce hit spread. The optical readout resolution, currently limited to 300 nm due to distortions, is improved by involving 3D image deconvolution. As for reconstruction, integrating machine learning methods offers substantial potential for resolving ambiguities caused by crossing multiple tracks in 3D space.

To keep the accumulated track density at an analyzable level, the emulsion films will be replaced once per year. The implementation of an effective sweeper magnet to reduce the muon fluence in the FPF would be beneficial.
Studies are ongoing to assess the effectiveness of a possible sweeper magnet in the LHC tunnel after the LOS has left the magnet cryostats, but before it leaves the tunnel.

The emulsion film production and its readout will be conducted at facilities in Japan. The capacity of the film production facility~\cite{Rokujo:2024gqw} is 1200 m$^2$ per year. The Hyper Track Selector (HTS) system~\cite{Yoshimoto:2017ufm} can read out $\sim$0.5 m$^2$ per hour, or 1,000 m$^2$ per year. 
Recently, an upgraded HTS system, HTS2, became operational with about two times faster speed. Scanning FASER$\nu$ films with HTS2 is under test. 
Analysis methodologies dedicated to TeV neutrino interactions are currently being developed and tested in FASER$\nu$. These methods include momentum measurements using  multiple Coulomb scattering information, electromagnetic shower reconstruction, and machine learning algorithms for neutrino energy reconstruction. First estimates of the efficiencies/performance for flavor-specific neutrino interactions were obtained~\cite{FASER:2019dxq}, and work is ongoing to refine them.

FASER$\nu$2 has a clear and broad physics target, and the detector is based on a well-tested technology for tau neutrino and short-lived particle detection. 
The performance of FASER$\nu$2 is based on the experience of the FASER$\nu$ detector operating at the LHC. Some novel aspects of FASER$\nu$2, such as how the detector is assembled, have been studied in dedicated test beams with positive results. 
Further studies are being carried out to optimize the detector performance, the detector operational environment, and the installation scheme.

\section{FORMOSA 
\label{sec:wg9}}

The FPF provides an ideal location for a next-generation experiment to search for BSM particles that have an electrical charge that is a small fraction of that of the electron. Although the value of this fraction can vary over several orders of magnitude, we generically refer to these new states as ``millicharged'' particles (mCPs). Since these new fermions are typically not charged under QCD, and because their electromagnetic interactions are suppressed by a factor of $(Q/e)^2$, they are ``feebly'' interacting and naturally arise in many BSM scenarios that invoke dark or otherwise hidden sectors. For the same reason, experimental observation of mCPs requires a dedicated detector. 

As proposed in Ref.~\cite{Foroughi-Abari:2020qar}, FORMOSA will be a milliQan-type detector~\cite{Haas:2014dda, Ball:2016zrp} designed to search for mCPs at the FPF.  FORMOSA will be technically similar to what the milliQan Collaboration has installed in the PX56 drainage gallery near the CMS IP at LHC Point 5 for Run~3~\cite{milliQan:2021lne}, but with a significantly larger active area and a more optimal location with respect to the expected mCP flux. As discussed in \cref{sec:physics} and shown in \cref{fig:BSM_newparticles} (left), FORMOSA has the potential to significantly extend the search for mCPs over the broad range of masses from 10 MeV to 100 GeV.

\begin{figure}[tbp]
  \centering
   \includegraphics[width=0.6\textwidth]{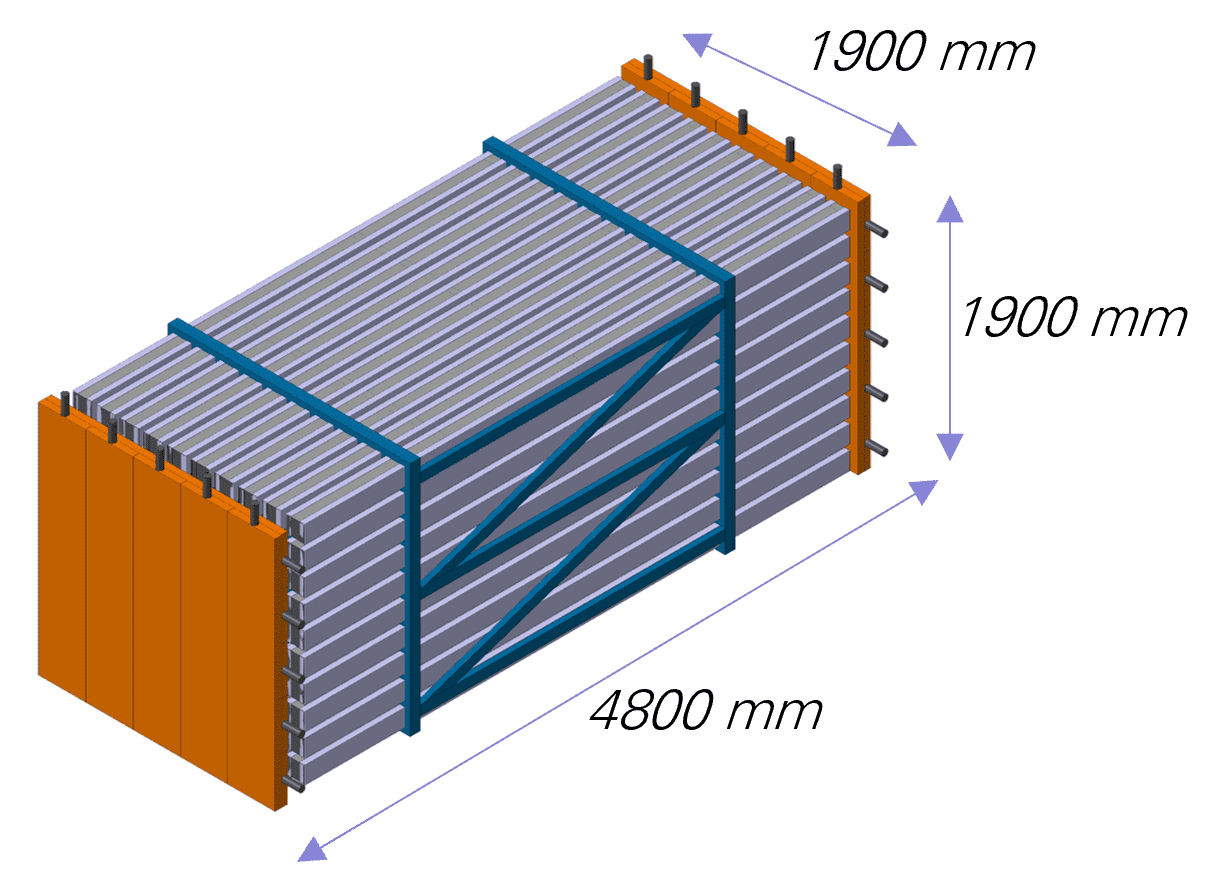}
  \includegraphics[width=0.38\textwidth]{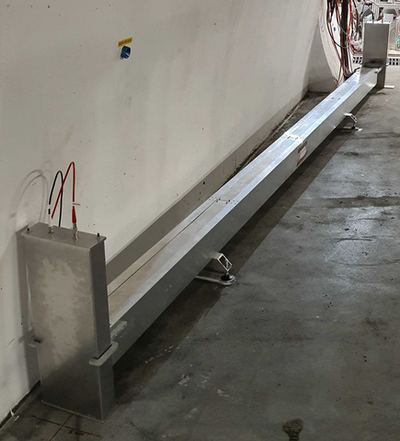}
\caption{Left: An engineering drawing of the FORMOSA detector. Right: The FORMOSA demonstrator taking data in the forward region of the ATLAS interaction point. }
  \label{fig:formosa-bars}
\end{figure}

To be sensitive to the small $dE/dx$ of a particle with $Q \lesssim 0.1e$, an mCP detector must contain a sufficient amount of sensitive material in the longitudinal direction pointing to the IP. As in Ref.~\cite{Haas:2014dda}, plastic scintillator is chosen as the detection medium with the best combination of photon yield per unit length, response time, and cost. Consequently, FORMOSA is planned to be a $1~\mathrm{m} \times 1~\mathrm{m} \times 5~\mathrm{m}$ array of suitable plastic scintillator (e.g., Eljen EJ-200~\cite{Eljen} or Saint-Gobain BC-408~\cite{SG}). The array will be oriented such that the long axis points at the ATLAS IP and will be located on the LOS. The array contains four longitudinal ``layers'' arranged to facilitate a 4-fold coincident signal for feebly-interacting particles originating from the ATLAS IP. Each layer in turn contains 400 $5~\mathrm{cm} \times 5~\mathrm{cm} \times 100~\mathrm{cm}$ scintillator ``bars'' in a $20\times20$ array. To maximize sensitivity to the smallest charges, each scintillator bar is coupled to a high-gain photomultiplier tube (PMT) capable of efficiently reconstructing the waveform produced by a single photoelectron (PE). To reduce random backgrounds, mCP signal candidates will be required to have a quadruple coincidence of hits with $\overline{N}_{\text{PE}} \ge 1$ within a 20 ns time window. The PMTs must therefore measure the timing of the scintillator photon pulse with a resolution of $\le5$ ns. The bars will be held in place by a steel frame. A conceptual design of the FORMOSA detector is shown in \cref{fig:formosa-bars} (left). 

Although omitted for clarity in \cref{fig:formosa-bars}, additional thin scintillator ``panels'' placed on each side of the detector will be used to actively veto cosmic muon shower and beam halo particles. Finally, the front and back of the detector will be comprised of segmented veto panels using perpendicular scintillator bars. This will provide efficient identification and tracking of the muons resulting from LHC proton collisions through the detector.  During Run 2 of the LHC, a similar experimental apparatus (the milliQan ``demonstrator'') was deployed in the PX56 draining gallery at LHC P5 near the CMS IP. This device was used successfully to search for mCPs, proving the feasibility of such a detector~\cite{Ball:2020dnx}.

Even though the pointing, 4-layered design will be very effective at reducing background processes, small residual contributions from sources of background that mimic the signal-like quadruple coincidence signature are expected. These include overlapping dark rate pulses, cosmic muon shower particles, and beam muon afterpulses. In Ref.~\cite{milliQan:2021lne}, data from the milliQan prototype was used to predict backgrounds from dark rate pulses and cosmic muon shower particles for a closely related detector design and location.  Based on these studies, such backgrounds are expected to be negligible for FORMOSA. Backgrounds from muon afterpulses are considered in Ref.~\cite{Foroughi-Abari:2020qar} and can be rejected by vetoing a 10 $\mu$s time window in the detector following through-going beam muons. This veto will be improved by the muon tracking provided by the segmented bars at the front and back of the detector. The feasibility of operating in the challenging forward region has been shown through the installation and operation of a prototype detector, the FORMOSA demonstrator, in the forward region of the ATLAS interaction point during Run 3 of the LHC. This is shown in \cref{fig:formosa-bars} (right). The FORMOSA demonstrator has taken data during Run 3 to validate the data acquisition strategy and measure backgrounds for the future FORMOSA detector.

The FORMOSA detector is proposed to be constructed of plastic scintillator, however, in the coming years, the exciting possibility of using alternative scintillator material with significantly higher light-yield will be studied. One such material is CeBr3 scintillator (available from Berkeley Nucleonics). This provides a light yield approximately factor 30 times higher than the same length of plastic scintillator with excellent timing resolution. This would allow much lower charges to be probed with the FORMOSA detector. 

\section{FLArE 
\label{sec:wg7}}

FLArE is a modularized, liquid argon, time-projection chamber (TPC) designed as a multi-purpose detector for a wide range of energies. It is motivated by the requirements of neutrino detection~\cite{Anchordoqui:2021ghd} and light dark matter searches~\cite{Batell:2021blf} and builds on the considerable investment in liquid noble gas detectors over the last decade (ICARUS, MicroBooNE,  SBND, ProtoDUNE, and various components of DUNE). In particular, the design of the FLArE detector has been informed by the design of the DUNE near detector~\cite{DUNE:2021tad} and the demonstrated performance of the ProtoDUNE detectors at CERN~\cite{DUNE:2021hwx}.  Liquid argon as an active medium allows one to precisely determine particle identification, track angle, and kinetic energy from tens of MeV to many hundreds of GeV, thus covering both dark matter scattering and high-energy LHC neutrinos. 

As a fully active detector with both ionization and scintillation capabilities, FLArE offers a unique scientific reach that complements other FPF detectors. In particular, its excellent timing resolution allows it to minimize event pile-up and reject muon-induced backgrounds. This permits lowering the energy thresholds down to approximately 30 MeV, therefore allowing searches for rare dark matter scatterings off electrons with typically low recoil energies~\cite{Batell:2021blf}. FLArE’s neutrino measurements will also significantly contribute to the neutrino physics program at the FPF. For example, its different detection technique generally provides an independent means of constraining systematic effects and backgrounds for neutrino measurements; the lower energy thresholds allow studies of neutrinos at lower energies compared to FASER$\nu$2; the strong rejection of muon-induced neutron backgrounds enables measurements of neutral current neutrino interactions to constrain NSI, the neutrino charge radius or the weak mixing angle; the large transverse size allows measurement of the neutrino flux over a wider rapidity range, extending the FPF's capabilities to constrain forward hadron production; the argon and iron targets, together with FASER$\nu$2 measurements on a tungsten target, will allow constraints on target-dependent effects, such as nuclear PDFs. Finally, the lack of pile-up also permits studies of muon DIS interactions, providing complementary data for structure function measurements and new physics searches

FLArE is expected to see about 25-50 high-energy neutrino events/ton/$\ifb$ of collisions, providing the opportunity to measure the neutrino fluxes and cross-section for all three flavors. The identification of tau neutrinos is a particularly challenging task requiring detailed simulations and reconstruction studies, but could in principle be achieved with high spatial and kinematic resolution. In addition, the large active volume and millimeter-level spatial resolution, along with excellent calorimetry, provide sensitivity to dark matter searches via electron scattering, as mentioned in \cref{sec:physics}.

\cref{fig:flare} shows the current baseline design for FLArE. A significant engineering effort has been carried out to define the detector geometry, cryogenics, and integration within the FPF cavern. The latest configuration is based on a single-wall $8.8~\mathrm{m} \times 2.0~\mathrm{m} \times 2.4~\mathrm{m}$ foam-insulated cryostat. The TPC is segmented in 21 modules, arranged in a $3 \times 7$ configuration. Each TPC module ($1.0~\mathrm{m} \times 0.6~\mathrm{m} \times 1.8~\mathrm{m}$) is divided into two volumes by a central cathode, with an anode at either end, resulting in 42 separate $30~\mathrm{cm}$ drift volumes. The modularity is needed for two main reasons: first, the muon rate at the FPF (\cref{fig:mu-flux}) is sufficiently high  that the space charge intensity requires a short gap ($<50~\cm$); and second, the trigger capability is enhanced by compartmentalizing the intense scintillation light from liquid argon. The total liquid argon fiducial (active) mass in this configuration is approximately $10~\mathrm{tons}$ ($30~\mathrm{tons}$). 

 \begin{figure}[tbp]
    \centering
    \includegraphics[width=0.45\textwidth]{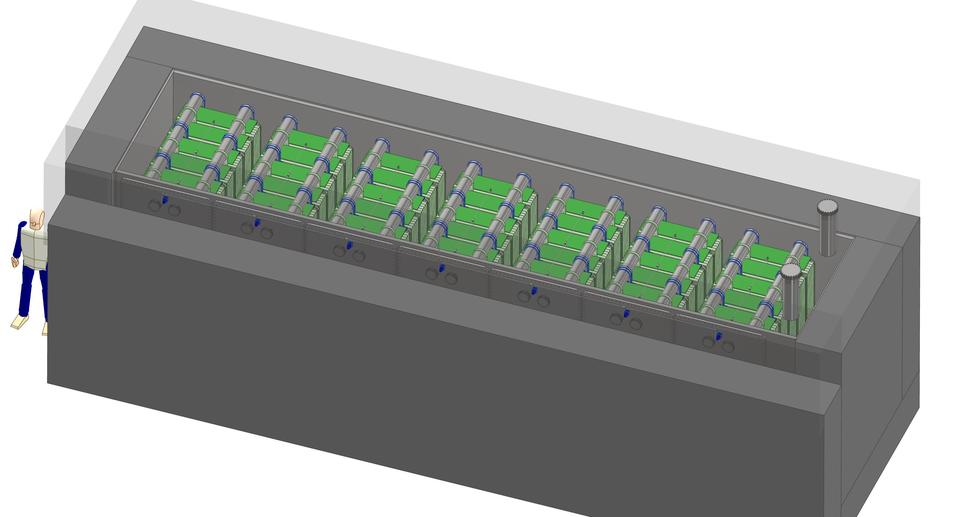}
    \includegraphics[width=0.45\textwidth]{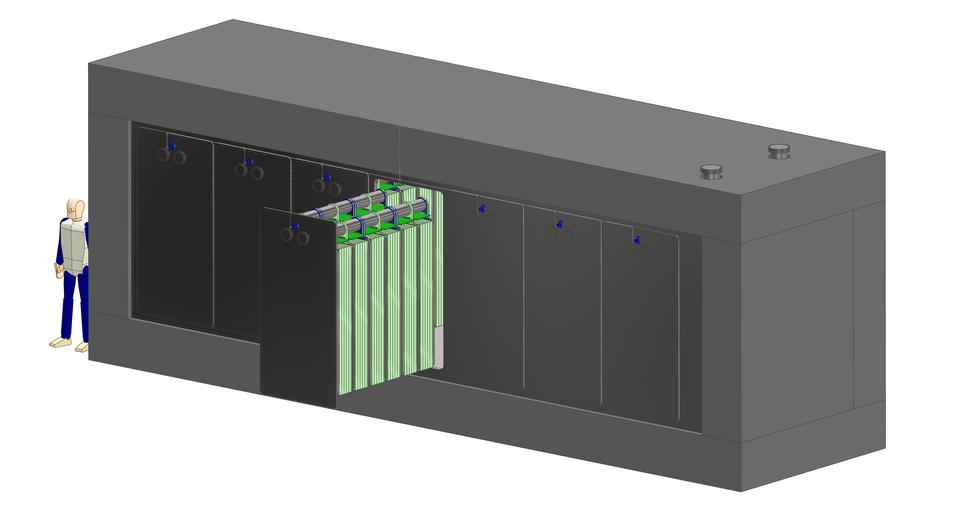}
    \caption{Layout of the FLArE baseline design. The detector is shown with the $3\times7$ modular segmentation. Three TPC modules are also shown  withdrawn horizontally from the cryostat.}
    \label{fig:flare}
\end{figure}

Given the limited height in the current design of the FPF cavern, the vertical insertion of the TPC modules into the cryostat is not possible. The insertion proceeds horizontally through doors on the side of the cryostat in a "filing cabinet" concept. A similar solution has been already successfully demonstrated in the EXO cryostat \cite{Auger:2012gs}. Each set of three TPC modules is mechanically supported via cantilevered beams by one of the cold doors. At the same time the door hosts the high-voltage feedthrough and flanges for readout electronics power and signal, as shown in \cref{fig:flare-tpc-module}. These assemblies can be easily transported into the cavern via wheeled carts. A custom machine holding the outer warm side of the door can then align and insert them, sealing the door against the cryostat itself. This procedure simplifies the installation and offers the possibility to extract single assemblies for maintenance or upgrades. 

Upon consultation with the CERN cryogenics experts, the cryogenic system for FLArE has gone through a substantial redesign. As shown in \cref{fig:facility-layout}, the downstream side of the FPF cavern is now reserved for some of the cryogenic infrastructure, including storage tanks for liquid argon, and nitrogen and a Turbo-Brayton LN2 condenser, which is a commercial unit that reduces the need to provide LN2 for cooling continuously. These facilities are kept away from the detectors to reduce noise and vibration.   The proximity cryogenics near the FLArE detector will simply consist of condensers and circulation systems for purity, all based on well known techniques from protoDUNE or ICARUS.  Briefly, LAr will be delivered at the surface and then transferred to the underground tank. The underground LAr storage tank will serve both as temporary storage and as an emergency cold vessel if the detector must be emptied quickly. LN2 will be delivered at the surface and filled in the Turbo-Brayton system to keep the detector and the LAr cold. 

The anode charge readout will be  pixelated. Preliminary simulations suggest that a $5~\mathrm{mm}$ pixel size will satisfy the spatial resolution requirements for track reconstruction and particle identification, as well as being reasonable from the point of view of electron diffusion, which diminishes the advantages of finer spacing. At a typical drift field of $500~\mathrm{V/cm}$, this translates to $\sim20,000$ electrons per pixel from minimum-ionizing muons and corresponds to a 30:1 signal-to-noise ratio assuming a total electronic noise of 500 electron equivalent noise charge (ENC). Concerning the electronics, two approaches are being considered: the LArPix ASIC~\cite{Dwyer:2018phu} developed for the DUNE Near Detector and the Q-Pix~\cite{Hoch:2024mce} readout scheme. Given the high number of pixels, 7200 per anode plane, careful considerations need to be taken to avoid an excessive heat load into the liquid. For instance, a proposed option to reduce the channel count consists of using a strip-based readout for the non-fiducial outer regions of the detector.

 \begin{figure}[tbp]
    \centering
    \includegraphics[width=0.45\textwidth]{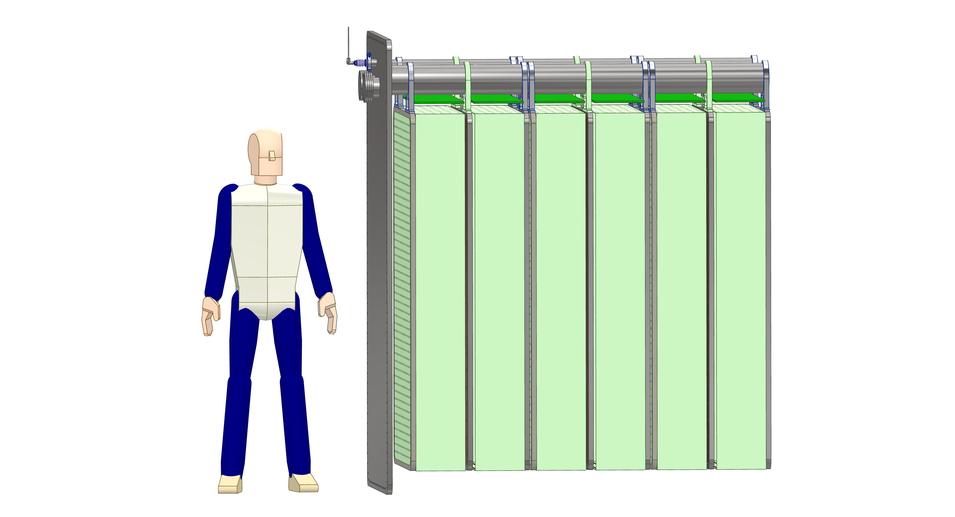}
    \includegraphics[width=0.45\textwidth]{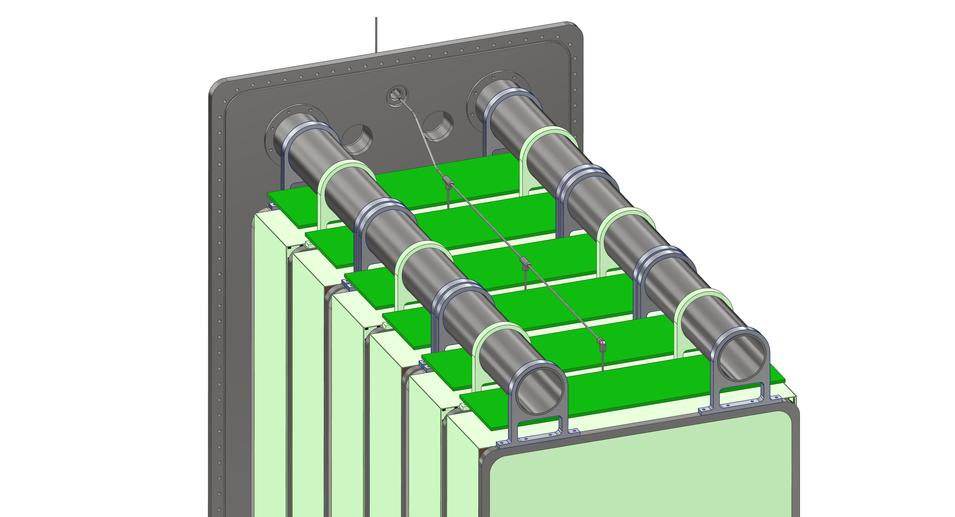}
    \caption{Left: TPC assembly with three TPC modules hanging from the cold door via cantilevered beams. Right: Inner view showing a conceptual high-voltage connection scheme to the cathode planes.}
    \label{fig:flare-tpc-module}
\end{figure}

An alternative readout design for FLArE is based on a 3D optical TPC similar to that developed within the ARIADNE programme. The ARIADNE approach utilises the 1.6~ns timing resolution and native 3D raw data of a Timepix3 camera to image the wavelength-shifted secondary scintillation light generated by a novel glass THGEM (THick Gaseous Electron Multiplier) within the gas phase of a dual-phase LArTPC~\cite{Roberts:2018sww, Lowe:2020wiq}. In this scenario, charge is drifted $1.8~\mathrm{m}$ vertically towards an extraction grid situated below the liquid level where they are transferred to the gas phase and subsequently amplified using a THGEM. The drift charge multiplication produces secondary scintillation light which is wavelength-shifted and imaged by Timepix3 cameras, providing a time sequence of 2D snapshots of the detector. This readout technology was successfully operated in a $2~\mathrm{m}\times2~\mathrm{m}$ prototype at the CERN Neutrino Platform~\cite{Lowe:2023pfk}. FLArE would be instrumented with 56 TimePix3 cameras, installed externally at cryostat view-ports. This design would lower the overall cost by eliminating the charge readout in favor of commercial and decoupled external devices, making it a valuable alternative to the more traditional TPC design.

No significant difference in physics performance is expected between the single-phase horizontal drift design and the dual-phase vertical drift design.  Diffusion is expected to be on the order of $1~\mathrm{mm}$ in both the transverse and longitudinal direction for both designs, and in both cases, timing depends on the strength of scintillation light.  Both options are therefore retained at this stage, and the decision between them will be based on cost and engineering considerations.

 \begin{figure}[tbp]
    \centering
    \includegraphics[width=0.6\textwidth]{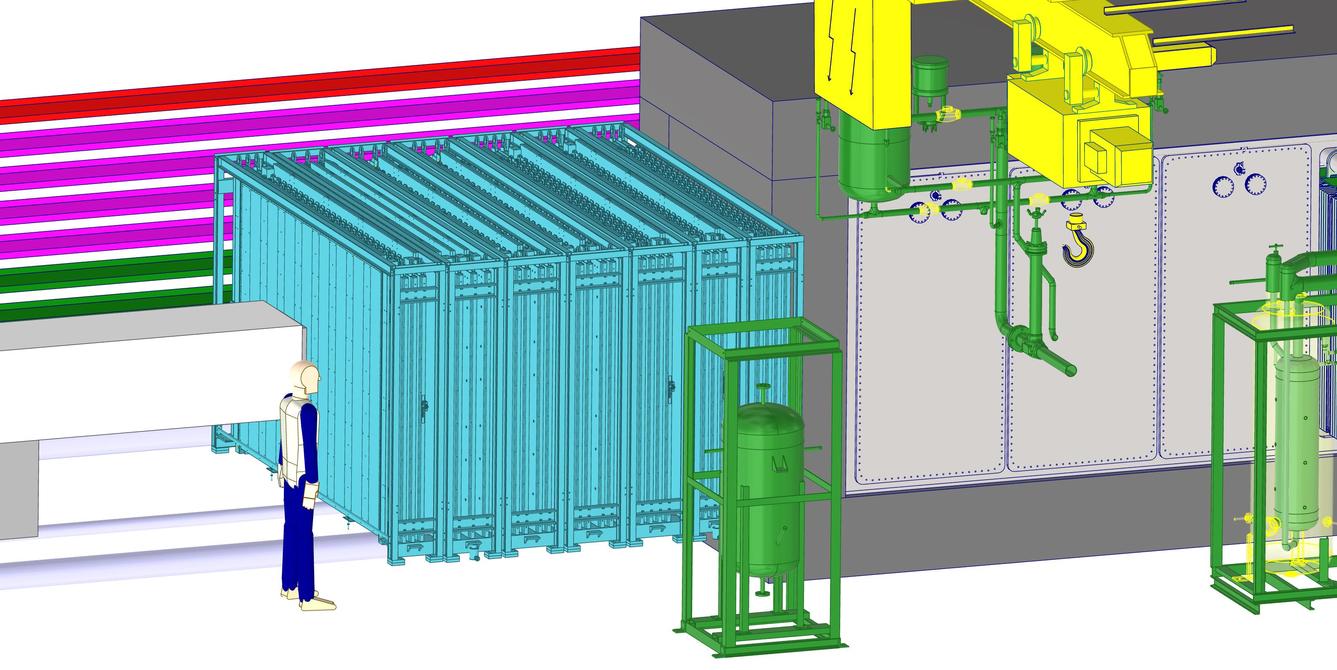}
    \caption{Preliminary design of the magnetized hadron/muon calorimeter downstream of the FLArE cryostat in the FPF cavern. The implementation is based on the Baby MIND concept \cite{Hallsjo:2018mmo}.}
    \label{fig:flare-baby-mind}
\end{figure}

One of the key requirements for FLArE is the ability to fully contain neutrino events and reconstruct their kinematics to identify the neutrino type. While the transverse size of the TPC ($1.8~\mathrm{m}$) was tuned with simulations for energy containment, energetic muons and a significant fraction of hadronic showers still escape the liquid argon volume along the line of sight. To improve energy containment and muon tagging, a magnetized hadron calorimeter/muon spectrometer is envisioned downstream of the TPC. Figure~\ref{fig:flare-baby-mind} shows a possible design based on the Baby MIND neutrino detector concept employed in the WAGASCI experiment~\cite{Hallsjo:2018mmo}. It consists of magnetized iron plates interleaved with scintillator modules that measure the particle position and the curvature of the track along the assembly. The clever magnetization scheme of the iron plates \cite{Rolando:2017rpx} allows one to achieve a $1.5~\mathrm{T}$ field inside the iron module with minimum stray field and operating current. This configuration avoids the need of bulky return yokes and cryogenic cooling, greatly easing its integration. Simulations are in progress to define the number and size of the plates for optimal containment and muon tagging efficiency, but a preliminary technical design has been adopted with a detector depth chosen to contain at least 90\% of energy from fiducial events. In addition, the synergy with the FASER2 magnetic spectrometer is also being investigated, since it is expected to provide up to $45-55\%$ acceptance for high-energy muons, depending on the final magnet design.

Overall FLArE will be an excellent neutrino detector that plays well with the FPF physics opportunities. Although additional R\&D is needed, the technical design is maturing quickly, and there is sufficient time and expertise available to complete all remaining tasks successfully within the FPF time frame.

\section{International Partnerships and Organization 
\label{sec:internationalparticipation}}

The FPF and the forward physics detector collaborations will follow the best governance practices  established by other major collaborations such as ATLAS, CMS, DUNE, etc.  The scale of the FPF enterprise is much smaller than these  collaborations, and so we will need to adjust the governance and coordination practices to be suitable.  We will also look at other models for this coordination, such as the LHCb or R\&D collaborations at CERN.  
An important difference between these major collaborations and the FPF community is the possibility of several independent FPF scientific collaborations using the same facility and sharing resources.  All of these collaborations will be international in nature.  

Three of the proposed experiments, FASER2, FASER$\nu$2, and FORMOSA, have pathfinder projects that are already installed and running at the LHC: FASER, FASER$\nu$, and milliQan, respectively. FASER has already placed world-leading bounds on dark photons~\cite{FASER:2023tle} and axion-like particles~\cite{FASER:2024bbl}, FASER$\nu$ has detected both electron and muon collider neutrinos~\cite{FASER:2023zcr} and measured their cross sections at TeV energies for the first time~\cite{FASER:2024hoe}, and milliQan has already placed world-leading bounds on mCPs~\cite{Ball:2020dnx}.  These results highlight the physics potential of the forward region, even with  modest luminosities and small experiments.  The collaborations behind the existing experiments are creating a community for this science that will serve all experiments in the FPF.  

Currently, the broader FPF experimental and theoretical community is approximately $\sim$400 strong; see the list of contributors to Ref.~\cite{Feng:2022inv}.  This community and additional scientists are expected to form the collaborations needed for the FPF experiments.   The community is currently dominated by scientists from Europe, US, and Japan.  Discussions to obtain R\&D support from their respective agencies and institutions are in progress.   

Along with the conceptual design report for the experiments and the facility, the community will propose a management   structure under the guidance of CERN. This structure will need to have a strong technical coordination team to construct the facility according to the scientific and engineering requirements and also to install the detectors. Each of the scientific collaborations will have representation in the technical coordination team and will provide scientific resources as needed. Each collaboration will also seek coordination and resolution of overlapping requirements. 

We expect the  scientific collaborations to  act independently to promote their respective  science and detectors, and seek collaborators from CERN and the broader international particle physics communities. CERN has a longstanding procedure for accommodating such new collaborators, mainly from Europe and North America.  For the FPF, we will attempt to broaden this to other parts of the world, especially in Asia, Africa,  and the Americas.  Such expansion will be welcome to inject new resources into this effort.  Fortunately, the current program for forward physics has created a core community that is centered around the experiments FASER, FASER$\nu$,  and milliQan.  This community is expected to lead the proposals for the FPF suite of experiments, but additional proposals that enhance the physics capabilities of the FPF are, of course, welcome.   

Below we make some remarks about the status of the collaborations for each of the constituent experiments along with what is needed for further expansion:  

\begin{itemize} 
\setlength\itemsep{-0.05in}
\item FASER2: The current FASER Collaboration has more than 110 members from 28 institutions in 11 countries~\cite{FASERWebpage}. FASER is a magnetized spectrometer (using permanent dipole magnets) that is housed in a service tunnel forward of ATLAS. It started taking data in Run 3.  The FASER Collaboration is expected to form the core of the FASER2 effort. FASER2 will require a much larger effort towards an appropriate spectrometer magnet and a larger tracking system that can handle the trigger rates from HL-LHC.  It also needs careful integration into the FPF hall and with the other experiments.  The Collaboration will need to expand to bring in the appropriate technical expertise and resources for the larger effort.   
\item FASER$\nu$2:  The FASER$\nu$2 collaboration will largely be made up from the existing FASER$\nu$ experts who are part of the 110-person FASER Collaboration. The expertise on emulsion is mostly concentrated in Japan, where there is deep expertise and a strong tradition in using emulsion-based detectors for neutrino physics. Japan has the leading facilities for emulsion gel production and for the scanning of the emulsion films after exposure, and these will both be used in the FASER$\nu$2 operations.  

\item FLArE: FLArE is based on the liquid argon technology developed for the FNAL short baseline program, as well as DUNE. The FLArE collaboration will be based on the current  working groups, which have approximately $\sim$50 participants, equally divided between US and European collaborators.  The collaboration has  received support from a private foundation, and a US national laboratory-  (BNL-)directed R\&D program.  Because of the recent investment in DUNE prototypes, only limited and well-targeted R\&D is needed for FLArE. Specifically, the readout electronics and pixel readout will need optimization for spatial resolution and dynamic range, however the majority of the design can be simply adapted from the DUNE ND-LAR design.  Furthermore, trigger strategies will need to be developed for the FLArE geometry.  At the moment, the collaboration has enough resources and person power to provide a physics proposal and a well-considered conceptual design.   A modest-sized international collaboration ($\sim$100 collaborators) with appropriate experience will have to be developed by the time of the technical design report in a few years.  

\item FORMOSA: FORMOSA would be based on the existing milliQan Collaboration with 29 members from 10 institutions in 5 countries~\cite{milliQanWebpage}.  The FORMOSA concept is based on  well-known technologies that require limited R\&D and is focused on mCPs.  Alterations and improvements to the design to substantially improve the detector sensitivity, such as through the use of alternative scintillator material, are under study, and the collaboration is expected to grow accordingly.

\end{itemize}  

\section{Schedule, Budget, and Technical Coordination 
\label{sec:schedule}}

A very preliminary budget and schedule is being  assembled for the FPF facility and the component experiments; this has been discussed in successive FPF workshops~\cite{FPF5Meeting, FPF6Meeting, FPF7Meeting}.  The costs are in several separate groups, as indicated in \cref{costtab}.  For this report we provide new cost numbers compared to the US-based estimates in Ref.~\cite{usp5fpf}: the civil construction design has been improved, providing more space and creating a section of the tunnel for cryogenic infrastructure away from the experiments~\cite{PBCnote2}, and the overall concepts for installation of all detectors and facilities have been improved. 

The cost for the civil construction and the outfitting was provided by the CERN civil engineering group and the technical infrastructure groups, respectively.  They reviewed the  initial experimental  requirements for the needed location, underground space, and services, and performed  a Class 4  estimate~\cite{CEcosting}.  According to international standards of conventional construction, a Class 4 estimate has a range of $-30\%$ to $+50\%$ around the point estimate.  The outfitting includes electrical service, as well as safety, ventilation, transportation, and lift services that are needed for the facility.  Obviously, the facility costs depend on the experimental requirements, which are expected to evolve as we progress towards a technical design. 

The costs for the experimental program were assembled by the proponents.  These  core costs are shown in \cref{costtab} for the FPF experiments.  The costs for FASER2 are dominated by the proposed magnetic spectrometer systems. For FASER$\nu$2, the costs are dominated by the production and handling of emulsion.  The international division of scope for components for these projects is currently not well defined, and therefore these core costs are provided without labor, overhead, contingency, and additional factors that must be used for a full cost estimate according to the rules of each national sponsor.  

FLArE and FORMOSA have substantial US portions; US cost estimates tend to include preliminary estimates for engineering, management, labor, overhead, and contingency factors.  These are not included in \cref{costtab} so that uniform core costs can be presented for each experimental project.  

FORMOSA is a conventional plastic scintillator-based detector with PMT readout.  The estimate includes mostly off-the-shelf parts and conventional assembly.  The number presented is for the more expensive commercial option for the readout  electronics. The FLArE estimate is based on the DUNE ND design with some modifications and includes a scintillator/steel hadron calorimeter using the Baby-MIND detector as a model~\cite{Hallsjo:2018mmo}.   The design will require targeted R\&D for the TPC electronics, a sophisticated photon sensor system, trigger electronics, and clean assembly. Granular details for the FLArE and FORMOSA costs including other factors are available at a pre-conceptual level.  These costs will be refined and further improved as we proceed to the conceptual design.

\begin{table}[t]
\begin{center}
\begin{tabular}{| p{0.27\linewidth} | p{0.17\linewidth} | p{0.51\linewidth} | }
 \hline \hline
\ Component &  \ Approximate Cost  &  \ Comments  \\  
 \hline
\ {\bf Facility Costs} &   &  \\ 
\ FPF civil construction &  \ 35.3 MCHF & \ Construction of shaft and cavern \\
\ FPF outfitting costs & \ 10.0  MCHF &  \ Electrical, safety, and integration \\ 

\ Cryogenic infrastructure  & \ 3.8 MCHF & \ Cryogen storage and cooling systems \\ 
\ {\bf Total} & \ {\bf 49.1  MCHF} & \ Includes integration for infrastructure \\ 
\hline 
\ {\bf Experiment Costs} & &  \ Core costs only\\ 
\ FASER2 & \  11.6 MCHF &  \ 3+3 tracker layers, SAMURAI-style  \\
 & & \ \ magnet, dual-readout calorimeter   \\ 
\ FASER$\nu$2 & \  15.9 MCHF & \ Tungsten target, scanning system, \\
 & & \ \ emulsion films (10 replacements), interface detector \\ 
\ FLArE  & \  10.8 MCHF & \ Cryostat,~proximity cryogenics, detectors  \\    
\ FORMOSA & \ 2.3 MCHF  & \ Plastic scintillator, PMTs, readout   \\ 
\ {\bf Total} & \ {\bf 40.6  MCHF} &  \ Core~cost~experimental program \\ 
\hline \hline
\end{tabular}
\end{center}
\vspace*{-0.15in}
\caption{Cost for components of the FPF and the experimental program.  Costs of the infrastructure at CERN are Class 4 estimates according to international standards; they have a range from $-30\%$ to $+50\%$.  
The costs for experimental components  are estimated as core costs, which consist of direct costs of materials and contracts only.  Each core cost was computed with conservative technical choices; as new ideas and designs are considered, the costs are expected to change.  
\label{costtab}}  
\end{table} 

A few additional comments are necessary for the project costs: 

\vspace*{-0.1in}
\begin{itemize} 
\setlength\itemsep{-0.05in}

\item The cost for FASER$\nu$2 includes the cost of replacing the emulsion films 10 times. These costs could change over time or be absorbed in the costs of detector operations.  

\item The baseline  for FLArE is now a single walled foam insulated 
cryostat that is opened on the side for installation. This design has been verified, but needs detailed reviews from laboratory experts.  The cost should be considered very preliminary; it represents a substantial savings compared to a membrane style cryostat.  The costs presented are for a pixel-style readout, but a very  significant option for FLArE is the ARIADNE optical readout option.  The cost of this option is dominated by the Timpix3 camera readout, but could be lower than the pixel option.   

\item Upon consultation with CERN experts, some of the cryogenic infrastructure has been separated from the experimental costs and included in the upper portion of the table.   The proximity cryogenics, which includes circulation and purification systems, is included in the FLArE experimental costs. 

\item Transport services will be needed for installation of large pieces such as the FASER2 magnet and the FLArE cryostat.  In addition, services will be needed for transporting the emulsion detector periodically.  
The cryostat/cryogenics and additional infrastructure design and costs clearly need to be coordinated and shared with CERN. This process of coordination has started only recently.  

\item The cost for  experiments  does not include engineering, labor, project management, contingency, and   the research support that will be needed.  Obviously, for an effort of this size, considerable support will be needed by a collaboration for students, postdocs, travel, and R\&D.  This is not included in the table. We estimate the total size of the collaboration to range from 250 to 350 people with corresponding annual support from the national agencies. 
\end{itemize}

The cost estimate for the FPF and its experiments will be refined in successive stages and reviews, as normally done for large  acquisitions.  We expect the review process to be defined by CERN, as the host lab, and the leading funding organizations that will be involved, such as the UK-STFC, US DOE and NSF, and Japanese JSPS.  Clearly, additional steps are needed to better define the scope of the facility and the constituent physics experiments.  The FPF community will continue with its working group activities and the FPF workshops. Detailed simulation activities have commenced and have provided critical information on detector sizes and depths needed for good efficiency and energy containment for various types of neutrino interactions, as well as for sensitive searches for the many possible BSM scenarios.  The CERN accelerator and radiation protection groups have contributed immensely by providing detailed simulations of the muon rates.  These simulation activities will require appropriate levels of support to develop the detailed requirements needed for a conceptual design report.  

\cref{tab:profile} has the proposed approximate funding profile using the current understanding of the cost estimates for components.   In the following we provide the constraints used for assembling this funding profile: 

\newcolumntype{x}[1]{>{\raggedright\arraybackslash}p{#1}}
\begin{table}[t]
\scriptsize
\renewcommand{\arraystretch}{1.3}
\begin{tabular}{|>{\raggedright}p{1.7cm}|>{\raggedright}p{1.25cm}|>{\raggedright}p{1.25cm}|>{\raggedright}p{1.25cm}|>{\raggedright}p{1.25cm}|>{\raggedright}p{1.25cm}|>{\raggedright}p{1.25cm}|>{\raggedright}p{1.25cm}|>{\raggedright}p{1.25cm}|>{\raggedright}p{1.25cm}|>{\raggedright}p{1.25cm}p{0.001cm}|}
\hline\hline
\bf{Year}                                                    & 2024                         & 2025                        & 2026                       & 2027                                     & 2028                        & 2029                         & 2030                                  & 2031             & 2032                                     & 2033         &                                                        \\
\hline
\bf{(HL-)LHC Nominal Schedule}                               & Run 3                        & Run 3                       & Run~3 / LS~3                      & LS 3              & LS 3 & LS 3  & LS~3 / Run~4                                 & Run 4            & Run 4                                    & Run 4  &                                            \\
\hline
\bf{FPF  Milestones}                                         & Pre-CDR and physics proposal & R\&D and prototype detectors & CDR, long lead items, magnet & Start of civil construction, TDR for detectors & Detector construction start & Major equipment acq. & End of civil construction, Install services & Detector install & Detector commissioning, physics start & Physics running all detectors &                            \\
\hline

\bf{Experiment Core Costs (kCHF)} &                              & 154                         & 1275                       & 3473                                     & 7257                       & 11220                        & 9503                                 & 6978            & 741                                     &                                                                            & 
\\
\hline\hline
\end{tabular}

\caption{
Proposed funding profile for the FPF  experimental program using the core cost numbers  from \cref{costtab}. The infrastructure cost profile is being developed.     The approval and cost rules will be different for the different sponsors who are proposed to contribute to this overall profile. Nevertheless, for the purpose of this illustration, the profile is shown in as-spent funds in a single currency.      }
\label{tab:profile}
\end{table}

\begin{itemize} 
\setlength\itemsep{-0.05in}
\item \cref{tab:profile} includes some milestones and the nominal HL-LHC schedule.  Any FPF construction must be coordinated with the HL-LHC, so that the civil construction and demands on personnel do not interfere with LHC operations. 
\item The CERN radiation protection group has concluded that the FPF can be accessed during LHC operations with appropriate controls for radiation safety.  This will allow detector installation to proceed during Run 4.  
\item An important constraint is that detector construction, installation, and commissioning must happen before Run 4 concludes.  This is quite important for the scientific productivity of the FPF and organization of the FPF community.  
\item We assume that the funding profile for the CERN infrastructure will follow the appropriate profile to allow start of detector installation in the 2030-2031 time frame.  
\item Full detector construction funding is assumed to start in 2027. However, critical development, such as the FASER2 magnet systems and FLArE cryostat, may require funding ahead of this date. Planning and integration of the FPF program will require excellent technical coordination with leadership from the host laboratories.  
\end{itemize}

\acknowledgements

We gratefully acknowledge the invaluable support from the CERN Physics Beyond Colliders study group, who have contributed many technical studies related to the feasibility of the implementation of the FPF.
We are grateful to Roshan Mammen Abraham, Max Fieg, and Jinmian Li for providing data for discovery reach plots.  
JA and ST are supported by the National Science Centre, Poland, research grant No.~2021/42/E/ST2/00031. JA is also partially supported from the STER programme -- Internationalisation of Doctoral Schools by NAWA. 
The work of LAA is supported by the U.S. National Science Foundation grant PHY-2412679.
The work of AJB is funded in part through STFC grants ST/R002444/1 and ST/S000933/1.
The work of BB is supported by U.S.~Department of Energy grant DE–SC-0007914.
The work of JB, MVD, SL, MV, and WW is supported in part by Heising-Simons Foundation Grant 2022-3319.
The work of MC is supported in part by U.S.~Department of Energy grant DE-SC0009999. 
The work of JLF and TM is supported in part by U.S.~National Science Foundation grants PHY-2111427 and PHY-2210283. 
The work of JLF is supported in part by Simons Investigator Award \#376204, Heising-Simons Foundation Grants 2019-1179 and 2020-1840, and Simons Foundation Grant 623683.  
The work of CSH is supported in part by U.S.~Department of Energy Grant DE-SC0011726.
The work of FK is supported by the Deutsche Forschungsgemeinschaft under Germany's Excellence Strategy -- EXC 2121 Quantum Universe -- 390833306.
The work of JM is supported by the Royal Society grant URF\textbackslash R1\textbackslash201519 and STFC grant ST/W000512/1.
The work of JR is partially supported by NWO, the Dutch Research Council, and by the Netherlands eScience Center (NLeSC).
The work at BNL is under U.S.~Department of Energy contract No.~DE-SC-0012704.  

\appendix
\section{Civil Engineering Costs}
\label{sec:appendix}

\cref{tab:CE-costs} shows a detailed breakdown of the CE costs. This is for the point estimate of 35.3 MCHF.  As a Class 4 estimate, the range is from $-30\%$ to $+50\%$.  

\begin{figure}[th]
\centering
\includegraphics[trim=0.cm 3.5cm 0.cm 0.cm,clip,width=0.9\textwidth]
{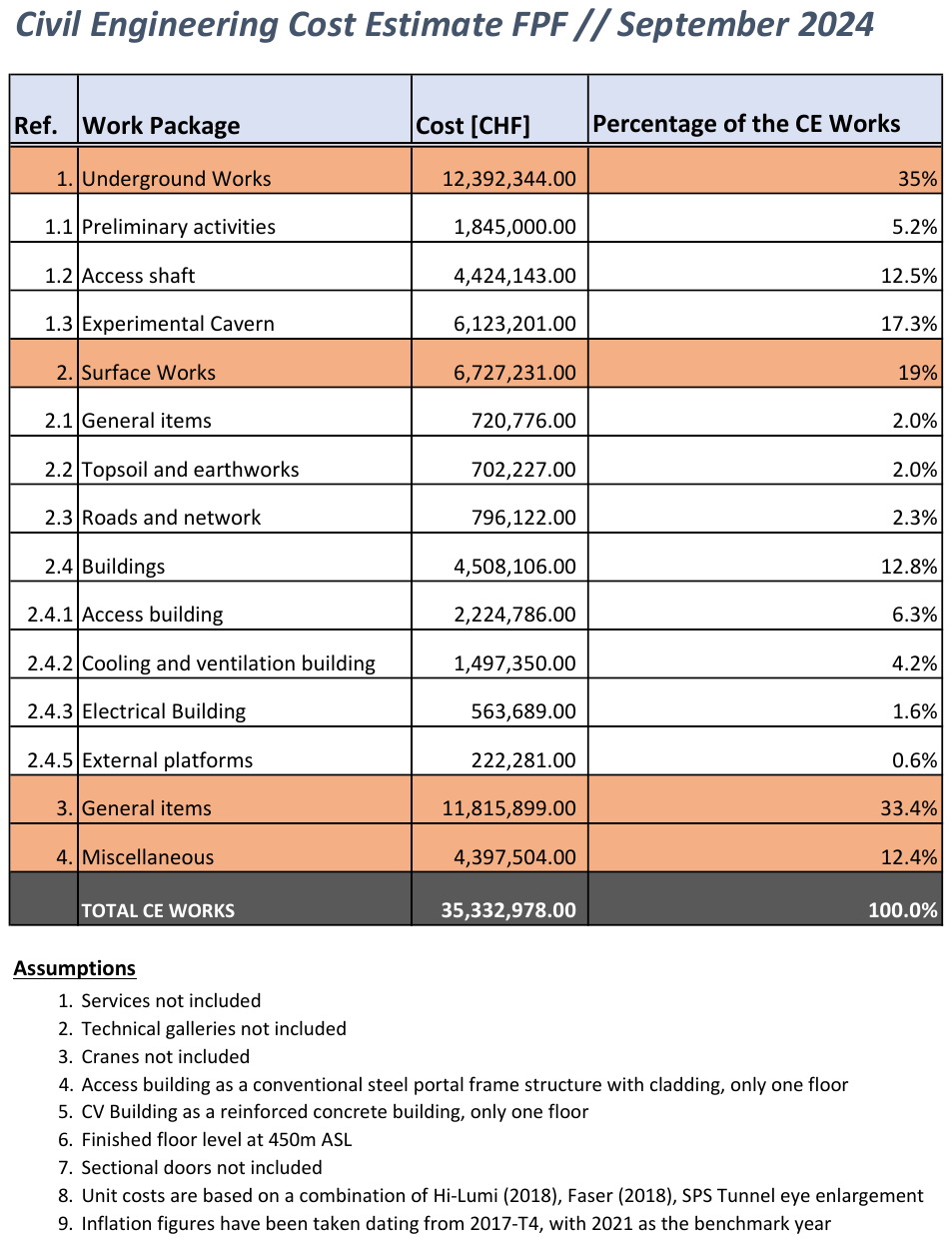}
\caption{A detailed breakdown of the costs of the baseline CE works. 
}
\label{tab:CE-costs}
\end{figure}
\newpage

\bibliography{FPF_Science_Planning}

\end{document}